\documentclass[anonymous=true]{article}
\PassOptionsToPackage{numbers, compress}{natbib}

\usepackage[preprint]{neurips_2023}

\usepackage[utf8]{inputenc} 
\usepackage[T1]{fontenc}    
\usepackage{hyperref}       
\usepackage{url}            
\usepackage{booktabs}       
\usepackage{amsfonts}       
\usepackage{nicefrac}       
\usepackage{microtype}      
\usepackage{bm}
\usepackage{tcolorbox}
\usepackage{marvosym}
\usepackage{xspace}
\usepackage{soul}
\usepackage[framemethod=tikz]{mdframed}
\usepackage{enumitem}
\usepackage{wrapfig}
\usepackage{caption}
\usepackage{multirow}
\usepackage{subfigure}
\usepackage{graphicx}
\usepackage{xcolor}
\usepackage{color}
\usepackage{colortbl}
\usepackage{amsmath}
\usepackage{amssymb}
\usepackage{pifont}
\usepackage{authblk}    
\newcommand{\ie}{\emph{i.e.}\xspace}
\newcommand{\eg}{\emph{e.g.}\xspace}
\newcommand{\aka}{\emph{a.k.a.}\xspace}

\newcommand{\wrt}{w.r.t.\xspace}

\newcommand{\cmark}{\ding{51}}%
\newcommand{\xmark}{\ding{55}}%
\definecolor{condition}{HTML}{E0EBF6}
\definecolor{green}{HTML}{54B345}
\definecolor{red1}{HTML}{B70404}

\definecolor{candidate}{HTML}{E5EFDB}

\definecolor{ginger}{rgb}{0.69, 0.4, 0.0}
\definecolor{pastelorange}{rgb}{1.0, 0.7, 0.28}
\definecolor{teal}{rgb}{0.0, 0.5, 0.5}
\definecolor{thistle}{rgb}{0.85, 0.75, 0.85}
\definecolor{mypink}{rgb}{.99,.91,.95}
\definecolor{mycyan}{gray}{0.9}
\definecolor{atomictangerine}{rgb}{1.0, 0.6, 0.4}
\definecolor{babyblueeyes}{rgb}{0.0, 0.5, 1.0}
\definecolor{aurometalsaurus}{rgb}{0.43, 0.5, 0.5}
\definecolor{mygray-bg}{gray}{0.9}
\newcommand{\name}{{Stream. CTR Prediction}}
\newcommand{\nname}{{Streaming CTR Prediction}}
\newcommand{\nnamee}{{Streaming CTR Prediction }}

\usepackage{makecell}
\setlist[itemize]{leftmargin=15pt}

\title{Streaming CTR Prediction: Rethinking Recommendation Task for Real-World Streaming Data}

\renewcommand\footnotemark{}

\author{%
  \textbf{Qi-Wei Wang}$^{1}$, \textbf{Hongyu Lu}$^{2}$, \textbf{Yu Chen}$^{2}$, \textbf{Da-Wei Zhou}$^{1}$, \textbf{\\De-Chuan Zhan}$^{1}$, \textbf{Ming Chen}$^{2}$, \textbf{Han-Jia Ye}$^{1}$
}
\affil{\small{$^{1}$State Key Laboratory for Novel Software Technology, Nanjing University\\ $^{2}$WeChat, Tencent}}

\begin{document}

\maketitle

\begin{abstract}

The Click-Through Rate (CTR) prediction task is critical in industrial recommender systems, where models are usually deployed on {\em dynamic streaming data} in practical applications. Such streaming data in real-world recommender systems face many challenges, such as distribution shift, temporal non-stationarity, and systematic biases, which bring difficulties to the training and utilizing of recommendation models.
However, most existing studies approach the CTR prediction as a classification task on {\em static} datasets, assuming that the train and test sets are independent and identically distributed (\aka, {\em i.i.d.} assumption). 
To bridge this gap, we formulate the CTR prediction problem in streaming scenarios as a {\bf \nnamee task}. Accordingly, we propose dedicated benchmark settings and metrics to evaluate and analyze the performance of the models in streaming data.
To better understand the differences compared to traditional CTR prediction tasks, we delve into the factors that may affect the model performance, such as parameter scale, normalization, regularization, etc. The results reveal the existence of the ``{\em streaming learning dilemma}'', whereby the same factor may have different effects on model performance in the static and streaming scenarios. 
Based on the findings, we propose two simple but inspiring methods (\ie, tuning key parameters and exemplar replay) that significantly improve the effectiveness of the CTR models in the new streaming scenario. 
We hope our work will inspire further research on streaming CTR prediction and help improve the robustness and adaptability of recommender systems.

\end{abstract}

\section{Introduction}

\begin{figure}[t]
    \subfigure[Performance Drop]{\includegraphics[width=0.48\textwidth]{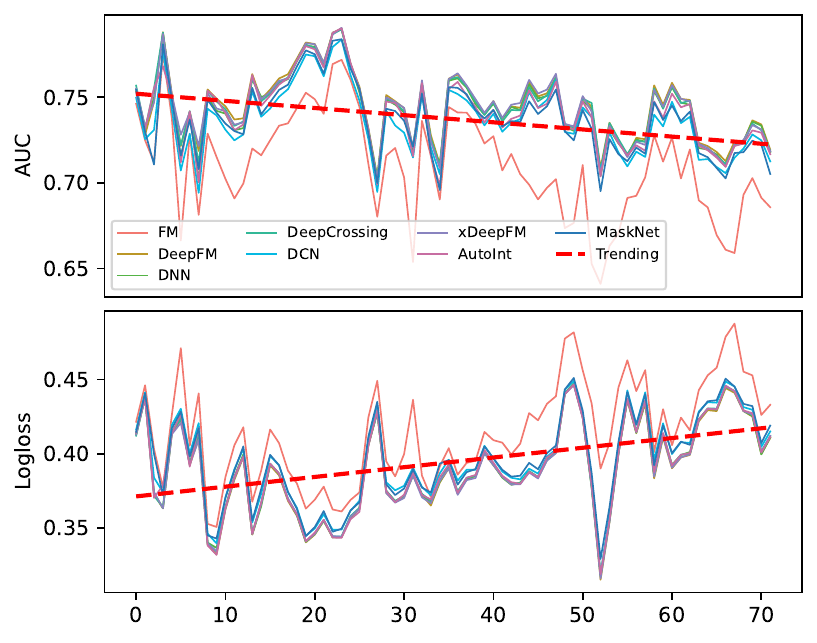}\label{fig:performance-drop}
    }
    \hspace{3mm}
    \subfigure[Distribution Shift]{\includegraphics[width=0.48\textwidth]{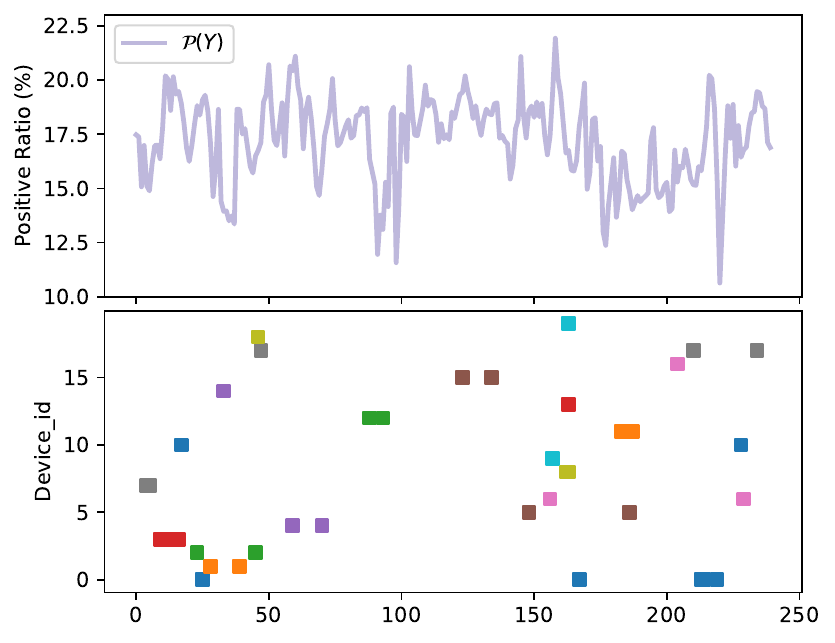}\label{fig:distribution-shift}
    }

    \caption{(a) An illustration of performance drop over time. We first average the performance (i.e., AUC and logloss) of all baseline methods and then fit a linear regression between the timestamps and the average performance to clarify the drop trending. (b) An illustration of the distribution shift from the {\em data view}. We first plot the positive ratio of different hours, which shows the $P(Y)$ usually drastically change over time. Besides, we randomly selected 20 device ids for the ``Device\_id'' field and displayed whether they appeared at different hours, which shows the $P(X)$ also drastically changes over time.}
    
\end{figure}

\label{sec:intro}

In an industrial recommender system, {\em data} naturally arrives in the form of a stream, which poses specific challenges such as label shifting~\cite{delayed-feedback-for-continuous-training,generalized-delay-feedback}, and systematic biases~\cite{bias-and-debias-survey,unbiased-position-aware,unbiased-learning-to-ran-feed-recommendation, previous-model-bias} to the updating of systems. The data stream is {\em time-aware} and usually contains significant {\em distribution shifts}. Therefore, the recommender system needs to be updated in a timely manner to ensure the model is up to date~\cite{streaming-rs}. The most widely-used approach to achieve this goal is incrementally updating the model when new data arrives, \ie, following a {\em streaming updating paradigm}. Some challenges in recommendation scenarios such as delayed feedback issue~\cite{delayed-feedback-for-continuous-training, generalized-delay-feedback} can be seen as derivative problems caused by the streaming training. Therefore, in this study, we concentrate on the {\em streaming training paradigm} and its impact on one of the most crucial tasks in recommender systems, namely predicting the Click-Through Rate (CTR) of items (\ie, whether a user clicks on an item).

Although real-world recommender systems are deployed in {\em streaming data} scenarios, the most critical task in recommender systems, \ie, the Click-Through Rate (CTR) Prediction task, is typically studied in {\em static} scenarios. Specifically, most existing studies~\cite{DNN_youtube,wide_deeplearning_google,DeepFM,DIN,deepcrossing,MaskNet,autoint,DCN,xdeepfm} usually treat the CTR Prediction task as a classification task (\ie, whether a user clicks on an item) on a {\em static and pre-collected} dataset. These studies typically partition the benchmark dataset (\eg, Avazu\footnote{\href{https://www.kaggle.com/c/avazu-ctr-prediction}{https://www.kaggle.com/c/avazu-ctr-prediction}}) randomly, or roughly by time, into two parts: the training set and a \textit{static} test set. A deep CTR Prediction model is then built on the training set and validated on the hold-out test set. The random partitioning assumes that the training and test sets are independently and identically distributed ({\em i.i.d.} assumption). 

However, the {\em dynamic streaming data} in real-world recommendation scenarios may exhibit significant distribution shifts between the train and test sets, violating the {\em i.i.d.} assumption and rendering the data {\em non-i.i.d}. Such distribution shifts in streaming data inevitably lead to performance degradation of the conventional CTR prediction model. To demonstrate this, we split the widely-used dataset (\eg, Avazu dataset) into training and test sets with \textit{non-overlapping timestamps} and train a CTR prediction model on the train set. We then evaluate the model with respect to each test timestamp. As shown in Figure~\ref{fig:performance-drop}, the test performance of various CTR prediction models all tends to degrade over time. To better illustrate the drop trending over time, we fit a linear regression model over the average performance (\ie, AUC and logloss) and timestamps and show the trend. The {\em decreasing} trend of AUC and the {\em increasing} trend of logloss over time strongly indicate that the performance of the conventional CTR prediction model is degraded over time, highlighting the importance of streaming updating the CTR prediction model to keep it up-to-date. Besides, we also show a significant distribution shift with respect to both the feature space $X$ and label space $Y$ in the Avazu dataset in Figure~\ref{fig:distribution-shift}. Specifically, we can observe that device ids (one of the features) appear intermittently in the streaming data process, without following any discernible pattern. This suggests that a {\em sudden drift}~\cite{Ensemble_and_streaming_learning} is ubiquitous in the streaming data of the recommender system.

Since {\em streaming data} exhibits significant challenges in recommendation systems, we aim to bridge the gap between {\em static} scenarios in existing studies and {\em dynamic} streaming scenarios in real-world applications in the context of CTR prediction. Despite there exist a few studies on incremental learning for CTR prediction~\cite{congcong_concept_drift,sigir23_asys,incCTR}, there still lacks a formal definition of the CTR prediction task in the streaming scenario and dedicated performance metrics. In addition, the target performance in streaming learning scenarios is not identical to that in incremental learning scenarios, so it is not straightforward to equate the two completely~\cite{gama2010knowledge}. Therefore, we formally formulate the {\bf \nnamee task} (\ie, \name) and propose corresponding performance metrics to evaluate the CTR prediction model in a streaming scenario. We decouple the \nnamee task into an inference under a distribution shift stage
({\em inference}) and a streaming learning stage ({\em updating}). In other words, the two most important characteristics of the \nnamee task are the two-stage paradigm of {\em inference $\rightarrow$ updating}  and the inference stage being conducted under a distribution shift scenario. 

To the best of our knowledge, there has been less work that has delved into the performance of existing deep CTR prediction models in streaming scenarios. Since the design and optimization of models are usually strongly correlated with the task setting, the lack of a streaming perspective may bias some studies from real-world applications.

To explore this issue, we propose a set of meaningful analysis metrics to investigate the impact of potential factors on the performance of the CTR prediction model in a streaming scenario. We empirically find that the model in static CTR prediction scenarios and streaming CTR prediction scenarios often exhibit {\em inconsistent preference} on certain factors, which is referred to as the ``{\em streaming training dilemma}'' in this study.

Furthermore, we tune the baseline methods with respect to the key factors identified from our empirical analysis and propose an exemplar-based method to improve online performance. Our contributions can be summarized as follows:
\begin{itemize}
    \item To the best of our knowledge, we are the first to formally formulate the CTR prediction task in the streaming scenario (\ie, { \nnamee task}) and propose corresponding performance metrics and analysis metrics to evaluate and analyze the model's performance.
    \item We are the first to explore and identify the key factors that tend to impact the performance of the deep CTR prediction model in the streaming scenario and empirically find the ``{\em streaming training dilemma}'' in the streaming scenario.
    \item We tune the CTR prediction models concerning the key factors identified from our analysis, which results in a {\em free-lunch} improvement on the performance in a streaming scenario without requiring extra effort. We also demonstrate that {\em exemplar replaying} can enhance CTR prediction models' performance in a streaming scenario.
\end{itemize}

\section{Related Work}
\label{sec:related}

\subsection{CTR Prediction}
\label{sec:related_ctr}
Click-Through Rate (CTR) prediction is a critical task in an industrial recommender system, which aims to predict whether a user will click the candidate item. Due to the strong feature interaction ability of Deep Neural Networks (DNNs), many DNN-based CTR prediction models~\cite{DNN_youtube,wide_deeplearning_google,DeepFM,DIN,deepcrossing,MaskNet,autoint,DCN,xdeepfm} are proposed to solve this task. Different from the structural image or text data, the data in CTR prediction is usually {\em sparse} and {\em unstructed}. In other words, the features in CTR prediction usually comprise different types. This high sparsity and the unstructured data type have led to the emergence of a joint embedding layer and fully connected layers to tackle this problem.

\subsection{Incremental Learning}
\label{sec:related_il}
Incremental learning aims to equip the model with continually learning new knowledge without forgetting old ones. Most existing studies of incremental learning only concentrate on image recognition~\cite{iCaRL,rmm,hou2019learning} or object detection task~\cite{DBLP:journals/pami/JosephRKKB22, DBLP:conf/iccv/WangWSG21} and only a few studies focus on incremental learning on recommender systems~\cite{congcong_concept_drift,sigir23_asys,session_based_memory_augmented_inc,session_based_exemplar_ader,incCTR}. Furthermore, only a limited amount of studies\cite{incCTR,sigir23_asys,congcong_concept_drift} concern the incremental learning of the CTR prediction model. These studies mainly focus on the catastrophic forgetting problem caused by concept drift and incremental learning. \cite{session_based_memory_augmented_inc,session_based_exemplar_ader} mainly focuses on the session-based recommendation system.

\subsection{Streaming Learning}
\label{sec:related_streaming}
Streaming learning aims to equip the model with learning ability in a {\em data stream}. The dynamic data stream poses new challenges for conventional machine learning models which are usually designed in a {\em static} environment. For example, the data stream is usually high-speed and accompanied by concept drift~\cite{Ensemble_and_streaming_learning,learning-under-concept-drift,the-problem-of-concept-drift} and evolvable features~\cite{feature-evolvable-stream-nips2017,evolvable-stream-aaai2021,evolvable-stream-icml2020}. \cite{Ensemble_and_streaming_learning} shows that ensemble learning is an effective technique to conquer the learning problem on a non-stationary data stream. Notably, incremental learning ability to conquer the data stream is necessary but insufficient~\cite{gama2010knowledge}. Besides, although there have been few studies discussing the challenges and evaluation methods~\cite{streaming_learning_evaluation,streaming_learning_RS_evaluation} of streaming training in recommender systems, most of them were proposed in the non-deep learning era and have become outdated.

\section{Formulation of Streaming CTR Prediction}

\begin{figure}[t]
  \centering
  \includegraphics[width=\linewidth]{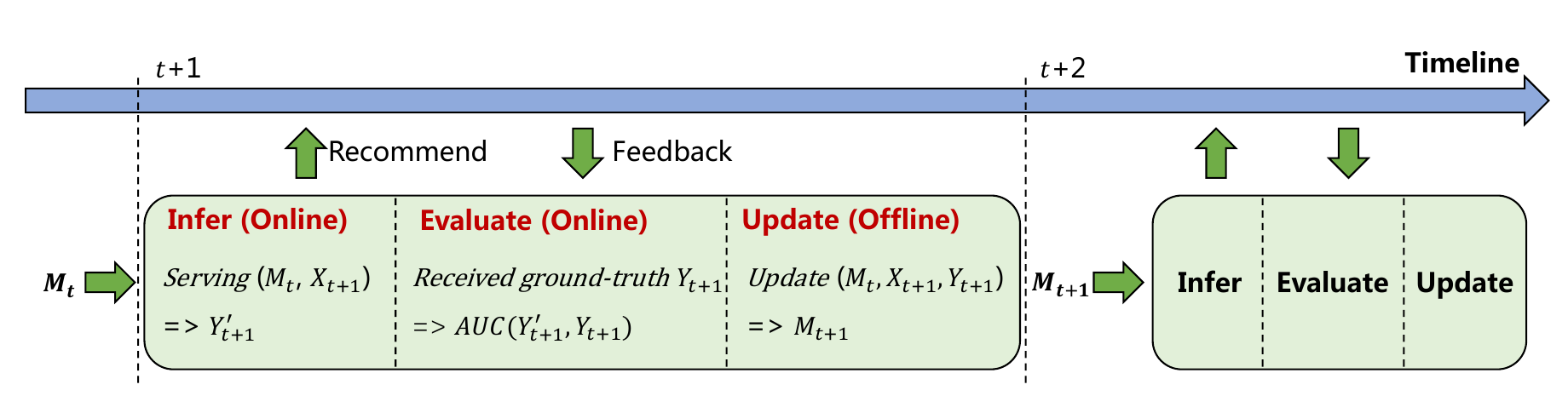}
  \caption{An illustration of recommender system in real-world streaming data. The streaming training follows {{\bf \em inference$\rightarrow$updating}} paradigm.}
  \label{fig:streaming-sys}
\end{figure}

\begin{figure}[t]
  \centering
  \includegraphics[width=\linewidth]{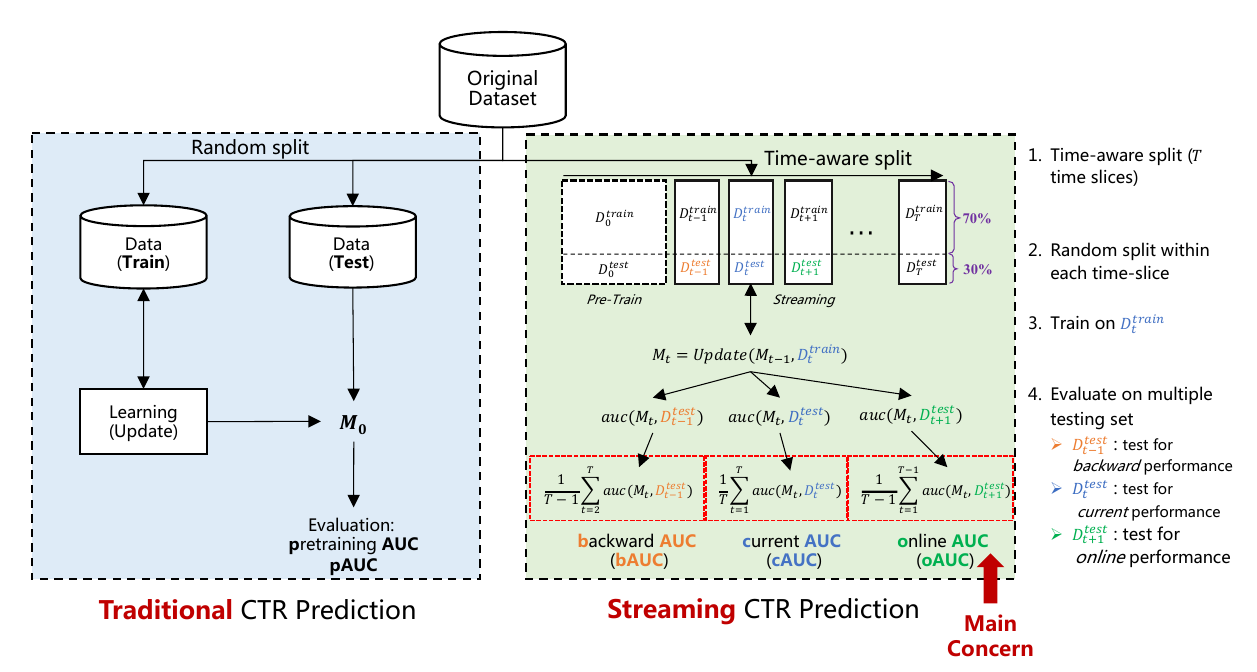}
  \caption{An illustration of the data splitting and performance metrics of the Streaming CTR prediction task. The notation ${\color{red}{auc}}(M_t,D_{t}^{test})$ means  compute the AUC of model $M_t$ on test set $D_{t}^{test}$.}
  \label{fig:streaming-ctr-data}
\end{figure}

\label{sec:streaming_ctr}
In this section, we first formally formulate the {\bf \nname} problem and introduce the detailed notations and the setting of benchmark datasets in section~\ref{sec:formulation} and section~\ref{sec:benchmark_setting}, respectively. We then propose new {\em \textcolor{red}{performance metric}} and {\em \textcolor{red}{analysis metrics}} for our empirical analysis in section~\ref{sec:online_auc} and section~\ref{sec:analy_metric}. Finally, we review the performance of existing CTR prediction models on the streaming scenarios based on the proposed performance and analysis metrics.

\subsection{Task Formulation}
\label{sec:formulation}
In this section, we give a formal formulation of the {\bf \nname} task. We expect the CTR prediction model could conduct streamingly infer and update in a data stream $\{D_t\}_{t=1}^{\infty}$. To make the formulation more concise, we assume the data stream with
fixed-length $T$ (\ie, $\{D_t\}_{t=1}^{T}$). Consider a specific timestamp $t$ and the corresponding training data $D_t^{train}$, the model $M_{t-1}$ and $M_t$ are {\em before and after} training the CTR prediction model on the training data of timestamp $t$, respectively. {\bf In the {\em inference} stage}, we expect the $M_t$ exactly predict the labels of $D_{\bm{t+1}}^{test}$. Then, the users {\em interact} with the recommended items in timestamp $t+1$ and generate the training data of timestamp $t+1$ (\ie, $D_{t+1}^{train}$). {\bf In the {\em updating} stage}, we update $M_t$ on $D_{t+1}^{train}$ and get the updated model $M_{t+1}$. The $M_{t+1}$ will be used to infer the incoming candidate sets $D_{t+2}^{test}$ and such a two-stage (\ie, {\em inference $\rightarrow$ updating}) training paradigm will periodically proceed with the arrival of streaming data. An intuitive streaming training pipeline is shown in Figure~\ref{fig:streaming-sys}.

Before this two-stage streaming learning paradigm, a warm-up {\bf \em pretraining} or batch-training stage on the data $D_0$ is conducted to get $M_{0}$. Notably, we further split the $D_0$ into nonoverlapping train set $D_0^{train}$ and test set $D_0^{test}$. A detailed illustration can be referred to in Figure~\ref{fig:streaming-ctr-data}.

To better understand the characteristics of the Streaming CTR Prediction task, we compare and summarize the differences between it and related tasks (\ie, CIL~\cite{CIL-survey}, TTA~\cite{TTA_survey} and CTR prediction~\cite{CTR_survey}). As shown in Table~\ref{tab:CIL-TTA-CTR}, the CIL task also deals with learning data streams, but it does not require time-correlated data or training only once (i.e., One Scan). By contrast, the TTA and CTR prediction do not involve a strict data stream. The TTA primarily addresses the feature shift problem during inference and does not require iterative training. The conventional CTR prediction task follows the {\em i.i.d} assumption and evaluates in the holdout test set. These distinctions highlight the unique characteristics of the \nname, which requires a formal definition and further exploration.

\begin{table}[]
\centering
\small
\caption{Differences between Class Incremental Learning (CIL), Test Time Adaption (TTA), CTR Prediction and Streaming CTR Prediction tasks from the {\em data} view. The ``label shift'' and ``feature shift'' indicate the presence of a shift between the training and test sets. ``One Scan'' refers to forwarding the training data through the model only once. ``Model in the loop'' refers to the training data of the current model is influenced by the stage of the previous model~\cite{previous-model-bias}.}
\renewcommand{\arraystretch}{1.3}
\resizebox{\textwidth}{!}
{\begin{tabular}{
>{\columncolor[HTML]{EFEFEF}}l |c|c|c|c|c|c}
\hline
 &
  \multicolumn{1}{c|}{\cellcolor[HTML]{EFEFEF}\textbf{Time-aware}} &
  \multicolumn{1}{c|}{\cellcolor[HTML]{EFEFEF}\textbf{label shift}} &
  \multicolumn{1}{c|}{\cellcolor[HTML]{EFEFEF}\textbf{feature shift}} &
  \multicolumn{1}{c|}{\cellcolor[HTML]{EFEFEF}\textbf{One Scan}} &
  \multicolumn{1}{c|}{\cellcolor[HTML]{EFEFEF}\textbf{Model in the loop}} &
  \multicolumn{1}{c}{\cellcolor[HTML]{EFEFEF}\textbf{Delayed Feedback}}
  \\ \hline
  
\textbf{CIL}& \xmark & \cmark & \xmark & \xmark & - & - \\
\textbf{TTA}& - & \xmark & \cmark & - & - & - \\ \hline
\textbf{CTR}& - & - & - & - & - & -  \\ 
\textbf{Stream. CTR}& \cmark & \cmark & \cmark & \cmark & \xmark & \xmark  \\ \hline
\end{tabular}}
\label{tab:CIL-TTA-CTR}
\end{table}

\subsection{Benchmark Setting}
\label{sec:benchmark_setting}
In this section, we give a detailed introduction to the benchmark setting of the Streaming CTR prediction task. We take the widely-used Avazu dataset as our benchmark dataset, which contains 10 days (240 hours) of labeled click-through data on mobile advertisements. We follow the data preprocessing in ~\cite{BARS,BarsCTR} of \textsl{Avazu\_x4\_001} and filter the infrequent features of the categorical fields by setting the minimum threshold as 2 for a fair comparison.

As shown in Figure~\ref{fig:streaming-ctr-data}, we take the first 70\% data (7 days, 168 hours) as {\em pretraining} data and the following 30\% data (3 days, 72 hours) as {\em streaming training} data. We {\em randomly shuffle} the pretraining data  and further split the 70\% pretraining data into a train set $D^{train}_{0}$ and left data as test set $D^{test}_{0}$. The procedure of training model $M_0$ on the train set $D^{train}_{0}$ and evaluating it on the test set $D^{test}_{0}$ can be seen as a {\em conventional} CTR prediction task and the $D^{train}_{0}$ and $D^{test}_{0}$ can be seen as independent and identically distributed. Besides, we also {\em randomly shuffle} the data of each hour in the streaming training stage and streamingly update the model $M_0$ on the subsequent data ($D^{train}_{i}, i\in [1, \cdots,T], T=72$) and evaluate the model on the hold-out test set ($D^{test}_{i}, i\in [1, \cdots,T], T=72$) to get following performance and analysis metrics. 

\subsection{Performance Metric}
\label{sec:online_auc}

Considering a CTR prediction model $M_{t}$ in timestamp $t$, the most important indicator for industrial application is the AUC of $M_t$ on $D_{t+1}^{test}$. Specifically, we denote the online AUC of $M_{\color{red}{t}}$ on $D^{test}_{\color{red}{t+1}}$ as ${\rm oAUC}_{\color{red}{t}}$, which indicates whether the CTR prediction model recommends the most favorite items to corresponding users. The overall \underline{o}nline \underline{AUC} (oAUC) performance of the CTR prediction model can be further denoted as:
\begin{equation}
    {\rm oAUC }= \frac{1}{T-1}\sum_{t=1}^{T-1} {\rm oAUC}_{t}
    \label{oauc}
\end{equation}
Notably, we only focus on the learning ability of the deep CTR prediction model and thus we neglect the performance of $M_0$ on $D_{1}^{test}$, which indicates the prediction ability of the pre-trained model $M_0$. Besides, the last model $M_T$ trained on $D_T^{train}$ has no following data and thus has no online performance (\ie, without ${\rm oAUC}_{T}$). {\em The oAUC can be seen as the \textcolor{red}{performance metric} that {\bf \em we are most concerned} within an industrial application.}

\subsection{Analysis Metrics}
\label{sec:analy_metric}
Although the oAUC can directly reflect the online recommendation quality, it only measures the model's performance in a {\em unified} way. Specifically, the oAUC is influenced by the pretraining stage and streaming training stage at the same time. Therefore, the oAUC can not indicate whether certain factor makes the pretrained model $M_0$ better suitable for the adaptation ({\em inference}) or make the model better {\em update} in a streaming scenario. Motivated by this, we propose some more 
\textcolor{red}{\em analysis metrics} for exploring and understanding the potential factors.

 First, we define some auxiliary analysis metrics except the performance metric \textcolor{green}{oAUC} (\ie, \textcolor{green}{online performance}). Since we have a hold-out test set for each timestamp (as seen in Figure~\ref{fig:streaming-ctr-data}), we can evaluate the model's performance in each test set to observe its performance trends. Specifically, we define the \underline{c}urrent \underline{AUC} (\textcolor{babyblueeyes}{cAUC}) and \underline{b}ackward \underline{AUC} (\textcolor{atomictangerine}{bAUC}). The computation of cAUC and bAUC is similar to oAUC and can be referred to the Figure~\ref{fig:streaming-ctr-data}. The cAUC and bAUC can be seen as an indicator to measure whether the CTR prediction model $M_t$ perform well on the current training distribution ($D_{t}^{test}$) and past data distribution ($D_{t-1}^{test}$). The bAUC can also be seen as an indicator to measure the {\em forgetting} degree of the model and the larger value of bAUC show the less forgetting of past knowledge. 

Notably, the oAUC, bAUC, and cAUC have only focused on the {\em streaming training} stage. Therefore, to provide a more comprehensive evaluation of the impact of corresponding factors on the pretrained model $M_0$ and the 
different preferable hyperparameters between the static and streaming scenario, we propose the metrics \underline{i}AUC and \underline{p}AUC to measure the effect of different factors only on the pretrained model  $M_0$. The main difference between these two types of AUC is the pAUC is the performance of $M_0$ on the hold-out pretraining test set $D_0^{test}$ and the iAUC is the average performance of $M_0$ on the $\{D_{2}^{test},\cdots, D_{T}^{test}\}$.
\begin{equation}
    {\rm iAUC} = \frac{1}{T-1} \sum_{t=2}^{T} auc({\color{red}{M_0}}, D_t^{test})
\end{equation}
Notably, we use iAUC to indicate whether some potential factors benefit the out-of-distribution generalization ability of the pretrained model $M_0$. We ignore the AUC of $M_0$ on the test set $D_1^{test}$ to align the test set between pAUC and oAUC. By contrast, pAUC measures the performance of $M_0$ on a traditional CTR prediction scenario.

After defining the performance metrics and analysis metrics, we first review the performance of the existing CTR prediction models in the scenario of static and streaming and show the results in Table~\ref{tab:CTRvsStreamCTR}. We annotate the ranking (\eg, {\scriptsize \colorbox{mycyan}{(1)}}) and variation (\ie, $\Delta$) of the same model in the static (\ie, the pAUC column) and streaming (\ie, the oAUC column) scenarios. Although most methods maintain a similar relative performance in terms of pAUC and oAUC, {\bf we observed that some CTR prediction models (\eg, FM and xDeepFM) exhibit significant differences in performance between these two scenarios}. This finding highlights the urgent need to closely examine and understand the behavior of CTR prediction models in streaming scenarios.

\begin{table*}[t]
\small
\caption{The Performance of different models for traditional CTR Prediction and Steaming CTR Prediction Tasks in the Avazu dataset. The $\Delta$ column shows the change in the ranking. The number (\eg, {\scriptsize \colorbox{mycyan}{(1)}}) following the pAUC and oAUC is the ranking of each model on the static and streaming scenario, respectively. {\scriptsize \textcolor{red}{$\blacktriangledown$ 2}} means the ranking drops by two spots and {\scriptsize \textcolor{green}{$\blacktriangle$ 1}} means the ranking increase by one spot.}
\renewcommand{\arraystretch}{1.1}
\resizebox{\textwidth}{!}
{\begin{tabular}{l|ll|lllllll}
\toprule
             & \multicolumn{2}{c|}{\textbf{CTR (Pre-Train)}} & \multicolumn{7}{c}{\textbf{Streaming CTR}}                    \\ \hline
 &
  \multicolumn{1}{c}{pAUC} &
  \multicolumn{1}{c|}{Imp.} &
  \multicolumn{1}{c}{bAUC} &
  \multicolumn{1}{c}{Imp.} &
  \multicolumn{1}{c}{cAUC} &
  \multicolumn{1}{c}{Imp.} &
  \multicolumn{1}{c}{oAUC} &
  \multicolumn{1}{c}{$\Delta$} &
  \multicolumn{1}{c}{Imp.} \\ \hline
FM~\cite{fm}      &0.7861\colorbox{mycyan}{\tiny (5)}&0.0106 & 0.7811 & - & 0.7773 & - & 0.7516\colorbox{mycyan}{\tiny (8)}&\tiny \textcolor{red}{$\blacktriangledown$ 3}& - \\
\hline
DNN~\cite{DNN_youtube}          &0.7893\colorbox{mycyan}{\tiny (1)}&0.0138 & 0.7936 & 0.0125 & 0.7885 & 0.0112 & 0.7691\colorbox{mycyan}{\tiny (1)}&-& 0.0174 \\
DeepFM~\cite{DeepFM}       & 0.7893\colorbox{mycyan}{\tiny (1)}&0.0138 & 0.7934 & 0.0122 & 0.7883 & 0.0110 & 0.7689\colorbox{mycyan}{\tiny (2)}&\tiny \textcolor{red}{$\blacktriangledown$ 1}& 0.0173 \\ 
DeepCrossing~\cite{deepcrossing} & 0.7893\colorbox{mycyan}{\tiny (1)}&0.0138 & 0.7929 & 0.0118 & 0.7879 & 0.0106 & 0.7689\colorbox{mycyan}{\tiny (2)}&\tiny \textcolor{red}{$\blacktriangledown$ 1}& 0.0173  \\
DCN~\cite{DCN}          & 0.7861\colorbox{mycyan}{\tiny (5)}&0.0106 & 0.7890 & 0.0079 & 0.7833 & 0.0060 & 0.7668\colorbox{mycyan}{\tiny (6)}&\tiny \textcolor{red}{$\blacktriangledown$ 1}& 0.0152 \\
xDeepFM~\cite{xdeepfm}&0.7755\colorbox{mycyan}{\tiny (8)}&- & 0.7949 & 0.0138 & 0.7907 & 0.0134 & 0.7681\colorbox{mycyan}{\tiny (4)}&\tiny \textcolor{green}{$\blacktriangle$ 4}& 0.0165 \\
AutoInt~\cite{autoint}     & 0.7850\colorbox{mycyan}{\tiny (7)}&0.0095 & 0.7867 & 0.0056 & 0.7822 & 0.0048 & 0.7654\colorbox{mycyan}{\tiny (7)}&-& 0.0138
 \\
MaskNet~\cite{MaskNet} &0.7877\colorbox{mycyan}{\tiny (4)}&0.0122 & 0.7976 & 0.0165 & 0.7939 & 0.0166 & 0.7670\colorbox{mycyan}{\tiny (5)}&\tiny \textcolor{red}{$\blacktriangledown$ 1}& 0.0154\\ \bottomrule
\end{tabular}}
\label{tab:CTRvsStreamCTR}
\end{table*}

\section{Revisiting Influential Factors}
\label{sec:main_exp}
The previous inconsistent performance of FM and xDeepFM in stationary and streaming scenarios motivates us to review the influential factors on the CTR prediction model under a streaming scenario. We first introduce the base experimental protocol details in section~\ref{sec:protocol} and then introduce the potential factors in section~\ref{sec:factors_intro}. Specifically, we explore the effect of model-related factors in section~\ref{sec:model-factor} and optimization-related factors in section~\ref{sec:optim-factor}. For the experiment of each factor, we first introduce the specific experimental protocol and then the corresponding experimental results.

\subsection{Default Protocol}
\label{sec:protocol}
To focus solely on the critical factor that affects performance under distribution shift, we have only split the entire dataset into the pretraining stage (70\%, the first 7 days) and streaming training stage (30\%, the following 3 days). Due to the diversity of CTR prediction model structures, we select the most representative structure (\ie, Embedding Layer and MLP network) as our baseline structure and follow the implementation in~\cite{BARS,BarsCTR}. We adopt the default hyperparameters setting proposed in~\cite{BARS} and only tune the specific hyperparameter that was explored while leaving the remaining hyperparameters as their default values.

\subsection{Influential Factors}
\label{sec:factors_intro}
In this section, we introduce the factors selected in this study and briefly introduce the reason for each selection. The core idea guiding us to make these selections is {\em which common factors may influence the model's performance in a distribution shift scenario}. Based on this core idea, we decouple the underlying factors into model-related factors and optimization-related factors. We hope to find a strong correlation between changes in certain common factors and the model's {\bf online performance} (\ie, oAUC) under the streaming scenario. Finally, we hope to answer the following important yet unexplored question: 
\begin{center}
    {\em How do different factors contribute to the performance of CTR prediction models in streaming scenarios?}
\end{center}

To our knowledge, we are the first to 
systemically formulate the {\bf \nnamee task} and further explore the answer to the above significant question. We hope to take a small yet significant step in a new direction for the training of CTR prediction models in the streaming scenarios.

\subsubsection{Model-related factors}
\label{sec:model-factors}
\begin{figure}[t]
  \centering \vspace{-5pt}
  \subfigure[Static CTR]{
  \includegraphics[width=0.4\linewidth]{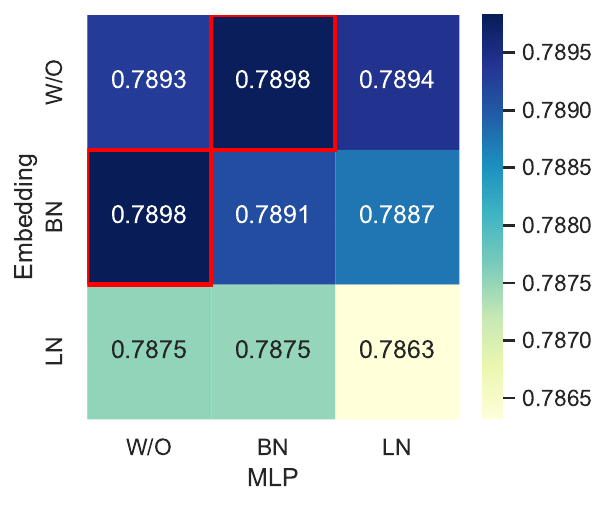}}
\subfigure[Streaming CTR]{
  \includegraphics[width=0.4\linewidth]{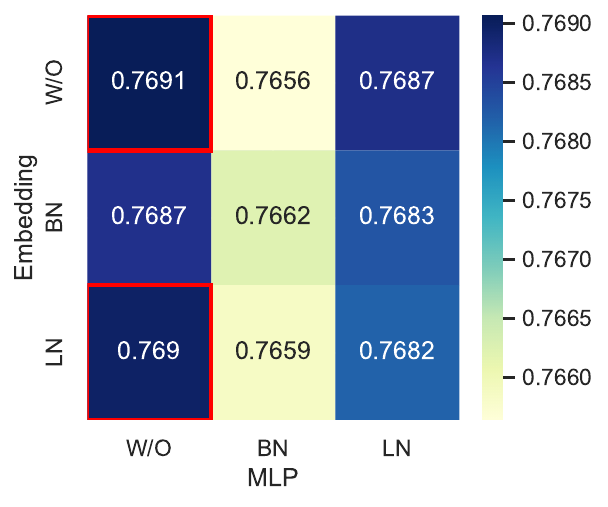}
  }
  \caption{The effect of normalization on the performance under the {\em static} and {\em Streaming} CTR prediction scenario. W/O means without normalization, BN means batch normalization and LN means layer normalization. The performance metric in the static scenario is pAUC, while in the streaming scenario, we use the proposed oAUC as the performance metric.}
  \vspace{-5pt}
  \label{fig:norm_ablation}
\end{figure}

\begin{figure}
    \centering
    \subfigure[Static CTR]{
    \includegraphics[width=0.8\linewidth]{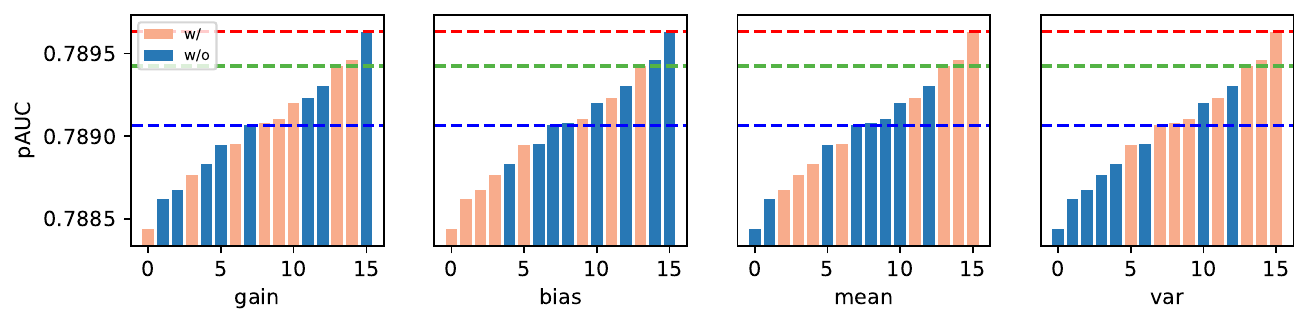}
    }
    \vspace{-3mm}
    \subfigure[Streaming CTR]{\includegraphics[width=0.8\linewidth]{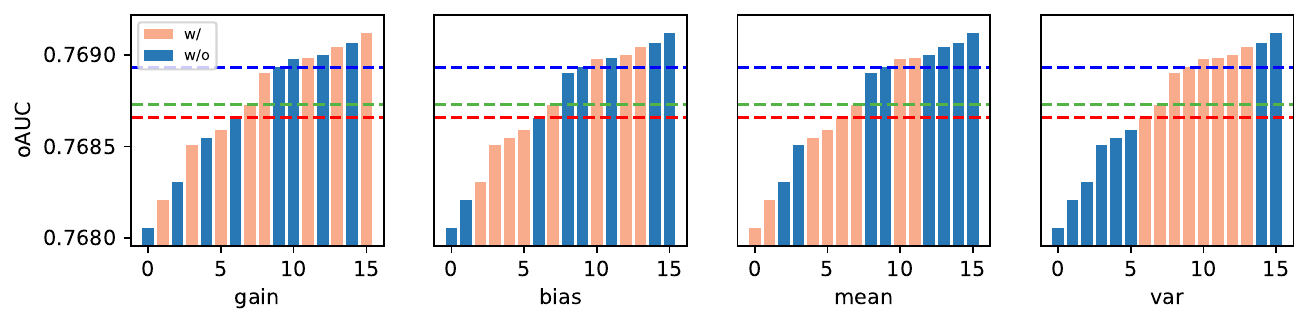}
    }
    \vspace{3mm}
    \caption{The {\bf LayerNorm} in {\em static} CTR prediction scenario and {\em streaming} CTR prediction scenario. We denote the SimpleLN in \textcolor{red}{red line}, the VO-LN in \textcolor{blue}{blue line} and the vanilla LN in \textcolor{green}{green line}.
    }
    \label{fig:LN_ablation}
\end{figure}

\begin{figure}
    \centering
    \subfigure[Static CTR]{
    \includegraphics[width=0.8\linewidth]{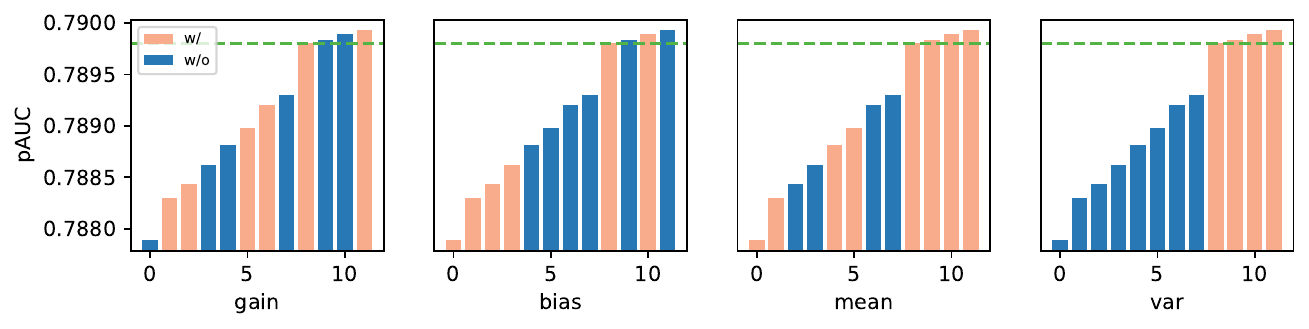}
    }
    \vspace{-3mm}
    \subfigure[Streaming CTR]{\includegraphics[width=0.8\linewidth]{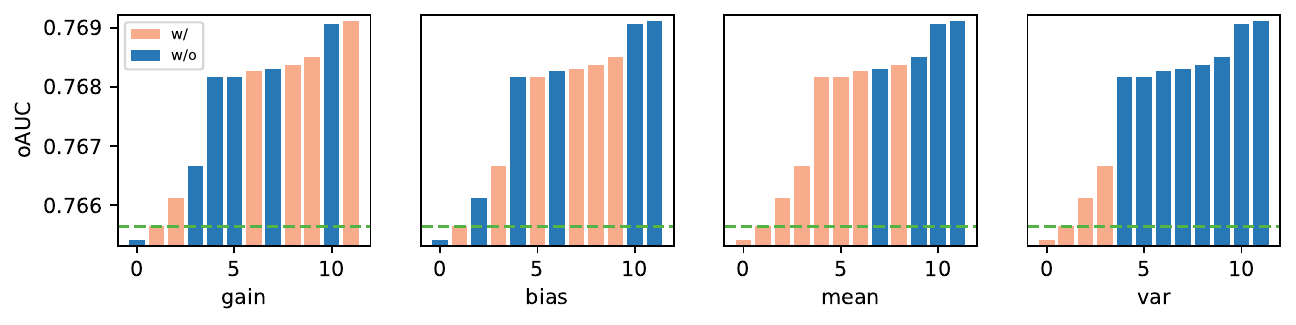}
    }
    \vspace{2mm}
    \caption{The {\bf BatchNorm} in {\em static} CTR prediction scenario and {\em streaming} CTR prediction scenario. We denote the vanilla BN in \textcolor{green}{green line}.}
    \label{fig:BN_ablation}
\end{figure}
\begin{itemize}
    \item {\bf Parameter size}: An Embedding layer followed by a Multilayer Perceptron (MLP) is the most widely used architecture in the CTR prediction task~\cite{DNN_youtube}. Due to the high sparsity of the user feature and item feature, the parameter size of the embedding layer is usually huge and dominates the total parameter size. The parameter size can be seen as a direct factor that leads the model ``{\em overfit}'' to the training distribution of the {\em current} timestamp and ``{\em underfit}'' the testing distribution of the {\em next} timestamp.
    
    \item {\bf Normalization}: Batch normalization~\cite{vanilla_BN} or Layer normalization~\cite{vanilla_LN} is widely used in DNNs to stabilize and accelerate training. However, the incorrect or careless usage of the normalization under a distribution shift scenario may lead to an enormous mistake~\cite{rethinking_batch,AdaBN}. The normalization under the distribution scenario is widely studied in computer vision and yet {\em not explored in the CTR prediction field}. Due to the inherent streaming characteristic of the recommender system, the exploration of normalization in the recommender system is also urgently needed. Notably, due to the special Embedding Layer and MLP structure in the CTR prediction model, there exist two types of normalization present: one in the embedding layer and the other in the MLP structure. 
    We analyze the influence of these two types of normalization on the CTR prediction model under a streaming scenario separately.
    \item {\bf Dropout rate}: The dropout~\cite{dropout} is a widely-used method to alleviate the overfitting phenomenon in the training of DNNs. Therefore, we explore whether the different degrees of dropout will influence the online performance of the deep CTR prediction model.
    \item {\bf Regularization}:
    The regularization item is widely attached to the loss function to restrict the optimization of model parameters and avoid the overfitting problem of DNNs~\cite{l2_regularization,PRML,ESL}. In the context of deep CTR prediction, regularization can be used to regularize the parameters of the embedding layer or the MLP network. In this study, we explore how the most widely used $L_2$ regularization of the embedding layer and MLP network  influence the CTR prediction model in a streaming scenario, respectively.
\end{itemize} 

\subsubsection{Optimization-related factors}
\begin{itemize}
    \item {\bf Optimizer}: The optimizer (\eg, Adam~\cite{adam}) is a key component to training a neural network successfully. Besides the default optimizer Adam, we explore whether the type of optimizer influences the performance of the model in a streaming scenario.
    \item {\bf Batch size}: The batch size is one of the most important hyper-parameters in the training of DNNs~\cite{concurrent-adversarial-learning-for-large-batch-training,second-order-gradient,incre-batch-size,large-batch-training}. We explore how the batch size influence the training of the CTR prediction model in a streaming scenario.
    \item {\bf Training epochs}: The training epochs in an industrial streaming learning scenario are usually set to 1~\cite{oneEpoch}. A simple and intuitive reason for this setup is the data streaming in the industrial system has a high speed and the increasing number of training epochs is impractical and inefficient. Besides, existing studies~\cite{oneEpoch} shows the CTR prediction model usually tends to have a performance dropping in the second epoch due to the overfitting problem. A more comprehensive exploration of this intriguing one-epoch phenomenon can be referred to in ~\cite{oneEpoch}. Motivated by this, we take the number of training epochs as a potential factor to influence the online performance of the model under the streaming training scenario. 
\end{itemize}








\subsection{Effects of Model-Related Factors}
\label{sec:model-factor}

\subsubsection{Normalization}
\label{sec:norm}
In this study, we only concentrate on the most widely used Layer Normalization (LN) and Batch Normalization (BN). These two types of normalization all following paradigms:
\begin{equation}
    y = \alpha * \frac{x-\mu}{\sqrt{\sigma^2+\epsilon}} + \beta
\end{equation}
The $\mu$ and $\sigma^2$ are the mean and variance computed from the unnormalized data. The $\alpha$ and $\beta$ are two affine parameters optimized by gradient descent. The most significant difference between BN and LN is the computation of $\mu$ and $\sigma^2$. The $\mu$ and $\sigma^2$ of BN are computed from the {\em batch} dimension and LN are computed from the {\em instance} dimension. Notably, the BN in training and testing mode has different types of normalization statistics. Specifically, the ($\mu_{\mathcal{B}}$, $\sigma^{2}_{\mathcal{B}}$) used in the training stage is computed from batch data, and the ($\mu_{EMA}$, $\sigma^2_{EMA}$) used in the testing stage is computed from the moving average.

\begin{equation}
\begin{aligned}
    \mu_{EMA} & \gets \lambda \mu_{EMA} + (1-\lambda)\mu_{\mathcal{B}}, \ \
    \sigma_{EMA}^{2} \gets \lambda \sigma_{EMA}^2 + (1-\lambda) \sigma_{\mathcal{B}}^2
\end{aligned}
\end{equation}

Existing studies~\cite{AdaBN,rethinking_batch,four-things-should-know-bn} show the BN is susceptible especially when encountering a distribution shift scenario. However, existing studies mainly focus on the computer vision field. How the influence of normalization on the CTR prediction model is unexplored. Besides,~\cite{simpleLN} shows that the affine parameters ($\alpha$ and $\beta$) in LN may lead to the overfitting problem and ~\cite{correct_norm} proposes a variance-only layer normalization in the context of CTR prediction:
\begin{equation}
    y = \frac{x}{\sqrt{\sigma^2 + \epsilon}}
\end{equation}
However, no studies explore how these normalizations and corresponding parameters (\ie, $\mu,\sigma^2,\alpha,\beta$) influence the CTR prediction model in a streaming scenario. 

In this section, we first train the model in different normalization combinations. Specifically, we first select the normalization of embedding layer and MLP from \{without normalization, batch normalization, layer normalization\}, respectively. For each normalization combination, we first evaluate $M_0$ on the conventional CTR prediction scenario (\ie, $D_0^{test}$) and obtain its static CTR prediction performance (\ie, pAUC). Besides, we can further update $M_0$ in the streaming CTR prediction scenario (\ie, the test set $D_t^{test}$ in the streaming training stage) and observe its online performance (\ie, oAUC). The results of different normalization combinations on conventional and streaming CTR prediction scenarios can be seen in Figure~\ref{fig:norm_ablation}.

\textcolor{red1}{\noindent \bf Finding:} As shown in Figure~\ref{fig:norm_ablation}(a), the usage of normalization in conventional CTR prediction scenario do not show extreme preference. The slightly low performance of LN on embedding shows that the LN may be unsuitable for the embedding layer. By contrast, the normalization in the streaming scenario shows a clear trend. As shown in Figure~\ref{fig:norm_ablation}(b), the performance of the model under the streaming scenario is highly correlated with {\em the normalization of MLP}. {\bf \em The model in the streaming scenario tends to achieve the best performance without normalization and the lowest performance with BN}, which indicates that vanilla normalization may harm the performance of the model in the streaming scenario.

Recognizing the significant importance of the normalization type used in MLP, we will delve deeper into the impact of different normalization parameters on online performance. Specifically, we take a comprehensive ablation study on the $\{\mu,\sigma,\alpha,\beta\}$ in LN and BN. We also observe the effect of certain factors on the performance of static CTR prediction and streaming CTR prediction. Due to the complex interplay and difficult-to-interpret nature of the hyperparameters, we only conclude the effect of certain factors from the performance trending.

\textcolor{red1}{\noindent \bf Finding:} As shown in Figure~\ref{fig:LN_ablation}, we find that $\{\mu, \sigma\}$ in LN tend to {\em benefit} the performance on static CTR prediction models, but {\em degrade} the performance of streaming CTR prediction models. As for the BN in Figure~\ref{fig:BN_ablation}, we also observe the {\em consistent} performance trending. The consistent observation indicates that the $\{\mu, \sigma\}$ may carry the distribution-related knowledge and are unsuitable for the context of streaming CTR prediction scenario.





\subsubsection{Parameter Size}
\label{sec:param_size}
\begin{table}[] 
\centering
\begin{minipage}[c]{0.32\textwidth}
\centering
\small
\includegraphics[width=1.0\linewidth]{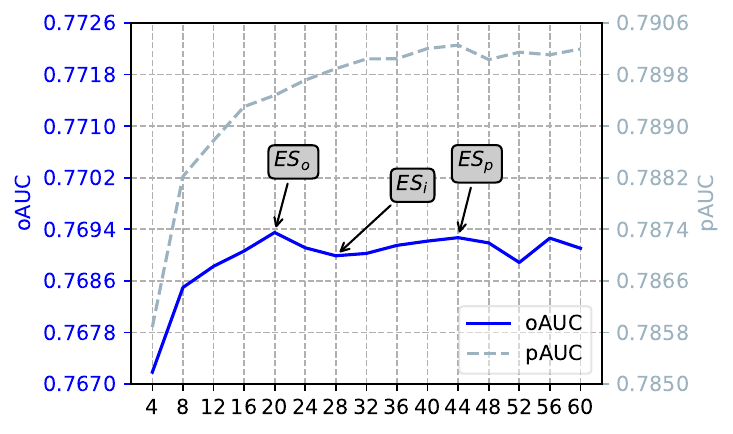}
\vspace{1mm}

\end{minipage}
\hfill
\begin{minipage}[c]{0.32\textwidth}
\centering
\small
\includegraphics[width=1.0\linewidth]{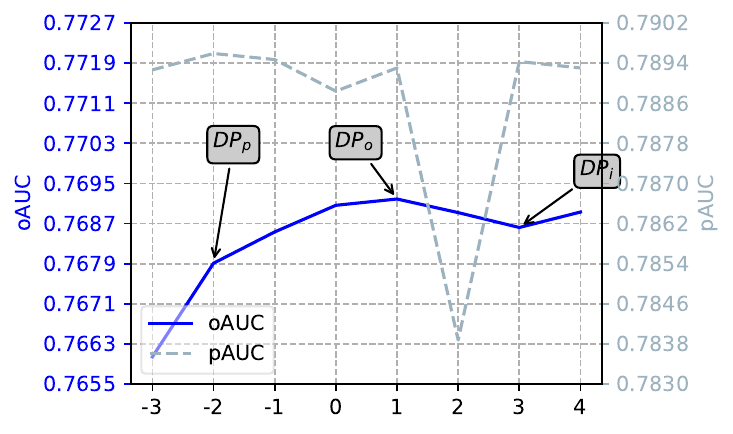}
\vspace{1mm}

\end{minipage}
\hfill
\begin{minipage}[c]{0.32\textwidth}
\centering
\small
\includegraphics[width=1.0\linewidth]{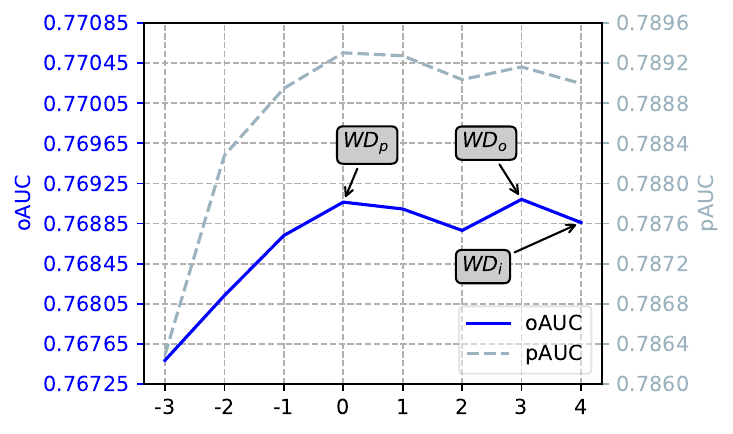}
\vspace{1mm}

\end{minipage}

\begin{minipage}[c]{0.32\textwidth}
\centering
\tabcolsep 1.5pt
\small
\vspace{-5pt}
\renewcommand{\arraystretch}{1.1}
\begin{tabular}{cccc}
    \hline
    & pAUC & oAUC & iAUC \\
    \hline
    \makecell{$ES_s$ \\ (4)} &0.7859 & 0.7672 & 0.7396 \\
    \hline
    \multirow{2}{*}{\makecell{$ES_m$ \\ (32)}}&0.7900 & 0.7690 & 0.7421 \\
      &\cellcolor{mycyan} \tiny {\bf +0.5290}
      &\cellcolor{mycyan} \tiny +0.2400 &\cellcolor{mycyan} \tiny +0.3300 \\
    \hline
    \multirow{2}{*}{\makecell{$ES_l$ \\ (60)}}&0.7902 & 0.7691 & 0.7417 \\
    &\cellcolor{mycyan} \tiny \textbf{+0.5490}
    &\cellcolor{mycyan} \tiny +0.2500
    &\cellcolor{mycyan} \tiny +0.2780 \\
    \hline
\end{tabular}
\caption{\small Emb. Size}
\vspace{-2mm}
\end{minipage}
\hfill
\begin{minipage}[c]{0.32\textwidth}
\centering
\small
\tabcolsep 4pt
\vspace{-5pt}
\renewcommand{\arraystretch}{1.1}
\begin{tabular}{cccc}
    \hline
    & pAUC & oAUC & iAUC \\
    \hline
    \makecell{$DP_{s}$ \\ (-3)}&0.7893 & 0.7660 & 0.7377  \\
    \hline
    \multirow{2}{*}{\makecell{$DP_{m}$ \\ (0)}}&0.7888 & 0.7691 & 0.7423 \\
      &\cellcolor{mycyan}\tiny -0.0540 &\cellcolor{mycyan}\tiny +0.3950
      &\cellcolor{mycyan}\tiny \textbf{+0.6250} \\
    \hline
    \multirow{2}{*}{\makecell{$DP_{l}$ \\ (4)}}&0.7893 & 0.7689 & 0.7420 \\
    &\cellcolor{mycyan}\tiny +0.0050
    &\cellcolor{mycyan}\tiny +0.3770
    &\cellcolor{mycyan}\tiny \textbf{+0.5890} \\
    \hline
\end{tabular}
\caption{\small Depth}
\vspace{-2mm}
\end{minipage}
\hfill
\begin{minipage}[c]{0.3\textwidth}
\centering
\small
\tabcolsep 4pt
\vspace{-5pt}
\renewcommand{\arraystretch}{1.1}
\begin{tabular}{cccc}
    \hline
    & pAUC & oAUC & iAUC \\
    \hline
    \makecell{$WD_{s}$ \\ (-3)} &0.7899 & 0.7695 & 0.7422 \\
    \hline
    \multirow{2}{*}{\makecell{$WD_{m}$ \\ (0)}}
      &0.7893 & 0.7691 & 0.7423 \\
      &\cellcolor{mycyan}\tiny +0.3830 &\cellcolor{mycyan}\tiny +0.2050 &\cellcolor{mycyan}\tiny \textbf{+0.6920} \\
    \hline
    \multirow{2}{*}{\makecell{$WD_{l}$ \\ (4)} }&0.7890 & 0.7689 & 0.74340 \\
    &\cellcolor{mycyan}\tiny +0.3440 &\cellcolor{mycyan}\tiny +0.1790 &\cellcolor{mycyan}\tiny \textbf{+0.8530} \\
    \hline
\end{tabular}
\caption{\small  Width}
\vspace{-10pt}
\end{minipage}
\caption{The effect of parameter size. Specifically, we analyze the effect of embedding size ({\bf ES}), the depth ({\bf DP}) and the width ({\bf WD}) of MLP. Setting the depth or width to 0 implies using a baseline structure configuration. A positive depth or width implies a deeper or wider network structure, while a negative one indicates a shallower or narrower one. For each pair of Figure and Table, the {\bf Figure} shows the trending of oAUC and pAUC about a certain factor and the {\bf Table} shows the detailed performance of pAUC/oAUC/iAUC on selected values. Specifically, $ES_p$ in Figure means the optimal {\bf E}mbedding {\bf S}ize (\ie, ES) for {\bf p}AUC (\ie, the subscript $p$). One column in the Table means the pAUC/oAUC/iAUC results of different embedding sizes. {\em Notably, we let the pAUC and oAUC have the same interval to make the trending of pAUC and oAUC in a unified figure meaningful}. For example, we select three embedding sizes in different scales (\ie, \underline{s}mall, \underline{m}edium and \underline{l}arge). The concrete value is shown in the Table. The values in \colorbox{mycyan}{gray shadow} mean the relative improvement (\%) compared to the baseline (the small scale). An example of computing the relative improvement of pAUC in medium scale can be referred to  $\frac{{\rm pAUC}\ \text{by}\ {\rm ES_m}\ -\ {\rm pAUC}\ \text{by}\ {\rm ES_s}}{{\rm pAUC}\ \text{by}\ {\rm ES_s}} \times 100\%$.} \label{fig:param_size}
\end{table}
\begin{table}[] 
\centering
\vspace{-3mm}
\begin{minipage}[c]{0.45\textwidth}
\centering
\small
\includegraphics[width=1.0\linewidth]{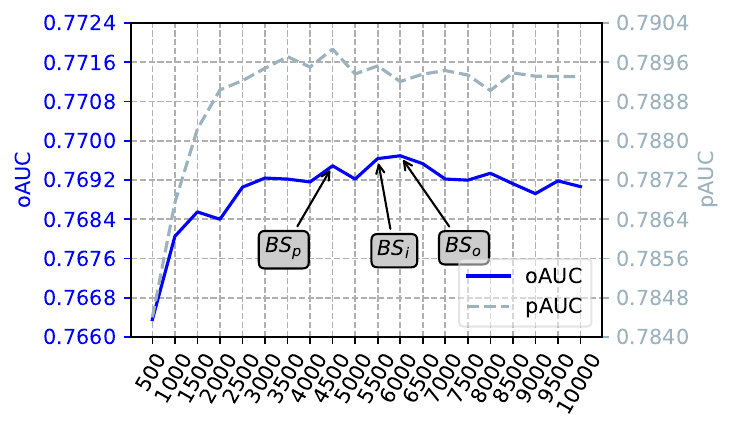}
\vspace{1mm}
\end{minipage}
\hfill
\begin{minipage}[c]{0.45\textwidth}
\centering
\small
\includegraphics[width=1.0\linewidth]{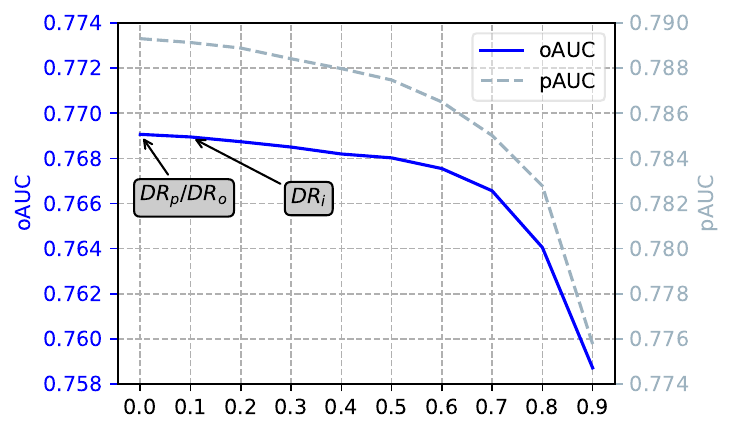}
\vspace{1mm}
\end{minipage}
\hfill

\begin{minipage}[c]{0.45\textwidth}
\centering
\tabcolsep 3.5pt
\small
\vspace{-10pt}
\renewcommand{\arraystretch}{1.1}
\begin{tabular}{cccc}
    \hline
    & pAUC & oAUC & iAUC \\
    \hline
    \makecell{$BS_s$ (500)}&0.7844 & 0.7664 & 0.7362 \\
    \hline
    \multirow{2}{*}{\makecell{$BS_m$ (5000)}}&0.7894 & 0.7692 & 0.7428 \\
      &\cellcolor{mycyan}\tiny + 0.632
      &\cellcolor{mycyan}\tiny + 0.374
      &\cellcolor{mycyan}\tiny {\bf + 0.902} \\
    \hline
    \multirow{2}{*}{\makecell{$BS_l$ (10000)}}&0.7893 & 0.7691 & 0.7423 \\
    &\cellcolor{mycyan}\tiny + 0.625
    &\cellcolor{mycyan}\tiny + 0.354
    &\cellcolor{mycyan}\tiny {\bf + 0.825} \\
    \hline
\end{tabular}
\caption{\small Batch Size}
\vspace{-2mm}
\label{fig_tab:bs}
\end{minipage}
\hfill
\begin{minipage}[c]{0.45\textwidth}
\centering
\small
\tabcolsep 3.5pt
\vspace{-10pt}
\renewcommand{\arraystretch}{1.1}
\begin{tabular}{cccc}
    \hline
    & pAUC & oAUC & iAUC \\
    \hline
    \makecell{$DR_{s}$ (0.0)}&0.7893 & 0.7691 & 0.7423 \\
    \hline
    \multirow{2}{*}{\makecell{$DR_{m}$ (0.5)}}&0.7875 & 0.7680 & 0.7406 \\
      &\cellcolor{mycyan}\tiny -0.2320
      &\cellcolor{mycyan}\tiny -0.1360
      &\cellcolor{mycyan}\tiny -0.2210 \\
    \hline
    \multirow{2}{*}{\makecell{$DR_{l}$ (0.9)}}&0.7758 & 0.7587 & 0.7358 \\
    &\cellcolor{mycyan}\tiny -1.7140
    &\cellcolor{mycyan}\tiny -1.3450 
    &\cellcolor{mycyan}\tiny -0.8720 \\
    \hline
\end{tabular}
\caption{\small Dropout Rate}
\vspace{-2mm}
\label{fig:dropout_rate_and_bs}
\end{minipage}
\hfill
\caption{The effect of Batch Size ({\bf BS}) and Dropout Rate ({\bf DR}). The interpretation of relative improvement and the abbreviations can be referred to in Table~\ref{fig:param_size}. The iAUC significantly benefits from the larger batch size.}
\vskip -10pt
\end{table}

\begin{table}[] 
\centering
\vspace{-3mm}
\begin{minipage}[c]{0.45\textwidth}
\centering
\small
\includegraphics[width=1.0\linewidth]{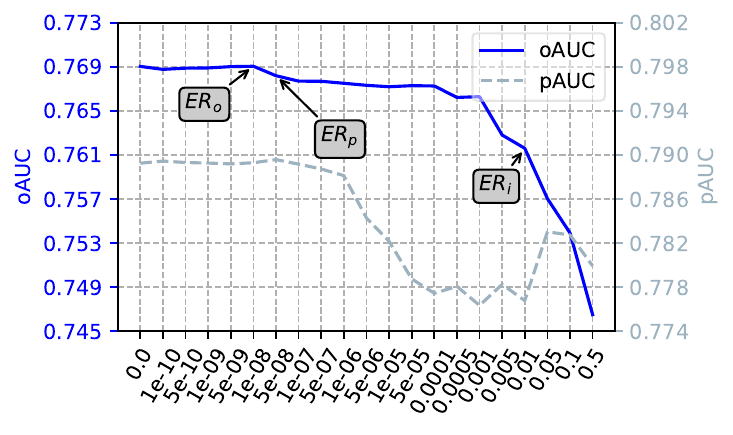}
\vspace{1mm}

\end{minipage}
\hfill
\begin{minipage}[c]{0.45\textwidth}
\centering
\small
\includegraphics[width=1.0\linewidth]{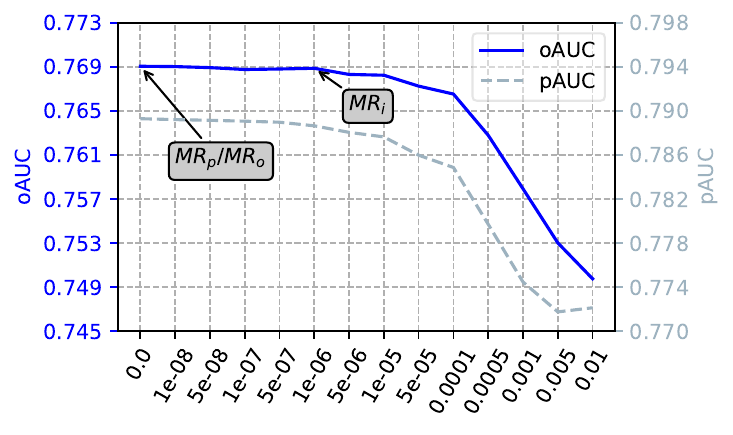}
\vspace{1mm}

\end{minipage}
\hfill


\begin{minipage}[c]{0.45\textwidth}
\centering
\tabcolsep 3.5pt
\small
\vspace{-10pt}
\renewcommand{\arraystretch}{1.1}
\begin{tabular}{cccc}
    \hline
    & pAUC & oAUC & iAUC \\
    \hline 
    \makecell{$ER_s$(0.0)} & 0.7893 & 0.7690 & 0.7428 \\
    \hline
    \multirow{2}{*}{\makecell{$ER_m$ (5e-6)}} &0.7843 & 0.7673 & 0.7473 \\
    
      &\cellcolor{mycyan} \tiny -0.6310
      &\cellcolor{mycyan} \tiny -0.2230 &\cellcolor{mycyan} \tiny {\bf +0.6010} \\
    \hline
     \multirow{2}{*}{\makecell{$ER_l$ (0.5)}} &0.7799 & 0.7465 & 0.7438 \\
    &\cellcolor{mycyan} \tiny -1.1820
    &\cellcolor{mycyan} \tiny -2.9310
    &\cellcolor{mycyan} \tiny {\bf +0.1280} \\
    \hline
\end{tabular}
\caption{\small Embedding Regularization}
\vspace{-2mm}
\end{minipage}
\hfill
\begin{minipage}[c]{0.45\textwidth}
\centering
\small
\tabcolsep 4pt
\vspace{-10pt}
\renewcommand{\arraystretch}{1.1}
\begin{tabular}{cccc}
    \hline
    & pAUC & oAUC & iAUC \\
    \hline
    \makecell{$MR_{s}$ (0.0)}&0.7893 & 0.7691 & 0.7423 \\
    \hline
    \multirow{2}{*}{\makecell{$MR_{m}$ (1e-05)}}&0.7876 & 0.7682 & 0.7415 \\
      &\cellcolor{mycyan}\tiny -0.2090 &\cellcolor{mycyan}\tiny -0.1060&\cellcolor{mycyan}\tiny -0.0990 \\
    \hline
    \multirow{2}{*}{\makecell{$MR_{l}$ (0.01)}}&0.7721 & 0.7498 & 0.7175 \\
    &\cellcolor{mycyan}\tiny -2.1750 &\cellcolor{mycyan}\tiny -2.5100 &\cellcolor{mycyan}\tiny -3.3370 \\
    \hline
\end{tabular}
\caption{\small MLP Regularization}
\vspace{-2mm}
\end{minipage}

\caption{The effect of Embedding regularization weight ({\bf ER}) and MLP regularization weight ({\bf MR}). The interpretation of relative improvement and the abbreviations can be referred to in Table~\ref{fig:param_size}. The trend of embedding layer regularization differs more significantly between the static and streaming scenarios. In contrast, MLP regularization shows a more consistent trend.}
\end{table}
\textcolor{red1}{\noindent \bf Finding:} We explore the parameter size of the CTR prediction model from three views, namely, the dimension of the embedding layer, the depth of the MLP network, and the width of the MLP network. We report the results in Figure~\ref{fig:param_size} and empirically observe the online performance (\ie, oAUC) is easier to saturate than the performance in static CTR prediction (\ie, pAUC). Besides the performance trend in the figure, the relative improvement of $ES_m/EM_l$ with respect to the embedding dimension in the table is also larger than oAUC and iAUC.

In addition, the oAUC with respect to \textbf{depth} shows a {\em increasing-then-saturation} trending and the pAUC does not follow this trend. By contrast, the effect of the \textbf{width} on oAUC and pAUC is consistent and oAUC achieved by $WD_o$ and $WD_p$ is similar. However, we empirically find that {\em iAUC significantly benefits from the increasing depth and width} compared to oAUC and pAUC, which indicates that increasing the parameter size of MLP may increase the out-of-distribution generalization ability of the model $M_0$.




\begin{figure}[!t]
    \centering
    \vskip -20pt
    \subfigure[Timestamp=0]{\includegraphics[width=0.32\textwidth]{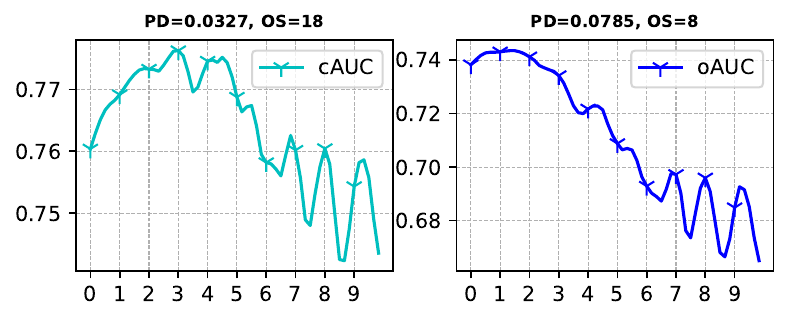}}
    \subfigure[Timestamp=1]{\includegraphics[width=0.32\textwidth]{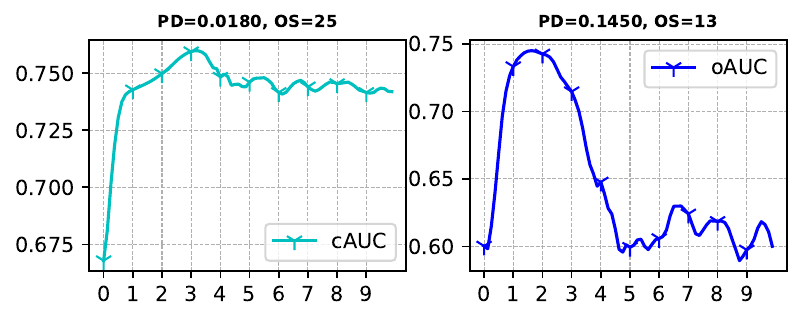}}
    \subfigure[Timestamp=15]{\includegraphics[width=0.32\textwidth]{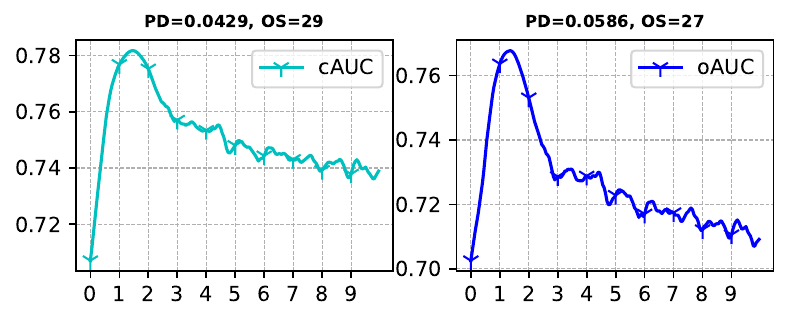}}
    \subfigure[Timestamp=33]{\includegraphics[width=0.32\textwidth]{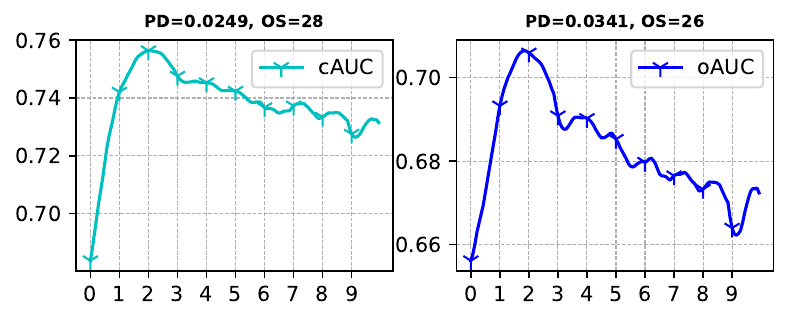}}
    \subfigure[Timestamp=51]{\includegraphics[width=0.32\textwidth]{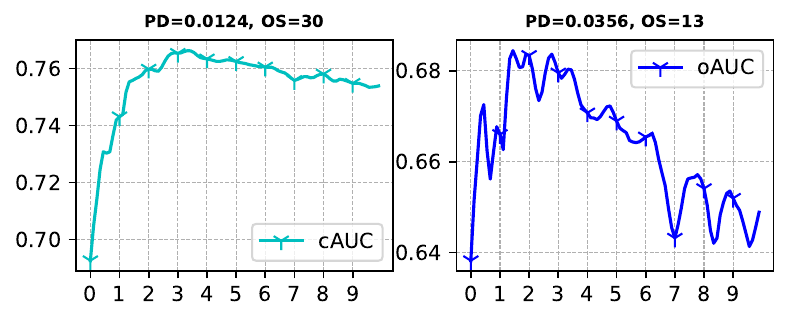}}
    \subfigure[Timestamp=69]{\includegraphics[width=0.32\textwidth]{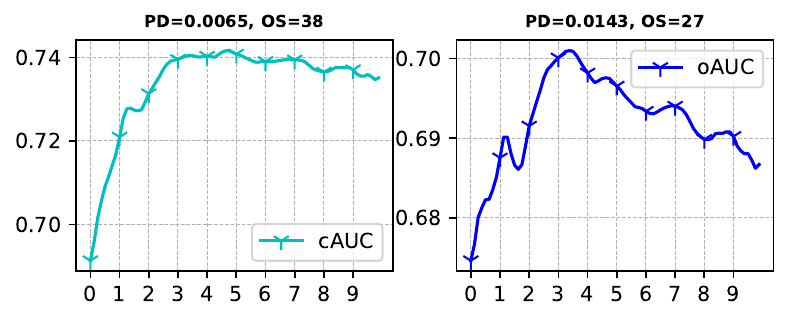}}
    \subfigure[The trending of optimal training steps and the performance drop.]{\includegraphics[width=0.8\textwidth]{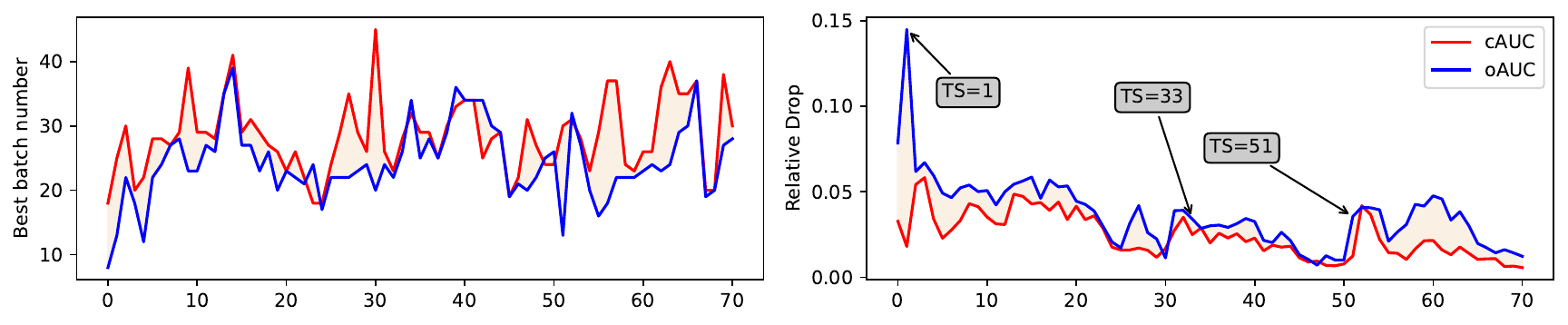}}
    \caption{We show the performance trending of oAUC and cAUC in (a)$\sim$(f). All the horizontal axis of figures (a)$\sim$(f) represent the index of training epochs. We also show the {\bf p}erformance {\bf d}rop ({\bf PD}) and the {\bf o}ptimal training {\bf s}tep ({\bf OS}) in each sub-figure. The performance drop means the relative drop between the best AUC and the last AUC. The optimal training step means the training steps to achieve the best performance \wrt oAUC or cAUC. The PD of oAUC is usually larger than the PD of cAUC, which means updating more on current distribution tends to make the oAUC drastically drop. Furthermore, the OS between oAUC and cAUC almost always differ, with the cAUC tending to be larger. This suggests that the optimal training step for cAUC may be suboptimal for oAUC.}
    \label{fig:epoch_numbers}
\end{figure}

\subsection{Effects of Optimization-Related Factors}
\label{sec:optim-factor}
\subsubsection{Optimizer}
In this section, we explore the effect of optimization-related factors on DNNs under the streaming  scenario and show the results in Table~\ref{tab:optim_analy}. We empirically show that Adam-based optimizers show relatively superior online performance.

\subsubsection{Batch Size}
\textcolor{red1}{\noindent \bf Finding:} As shown in Table~\ref{fig_tab:bs}, we observe that although the trend of pAUC and oAUC all follows the {\em increasing-then-saturation} paradigm, the optimal batch size for oAUC and pAUC may be different. This mismatch dilemma reminds us to pay closer attention to the online performance of the model, as some optimal factors for the performance in a static scenario may be suboptimal for the streaming scenario.


\subsubsection{Training Epochs}

We train for 10 epochs for each timestamp in the streaming data and have the following observations and findings from Figure~\ref{fig:epoch_numbers}:
\begin{itemize}
    \item {\bf The optimal steps for oAUC and cAUC are usually {\em mismatched}, especially at the beginning of streaming learning.} As shown in the (a)$\sim$(f) and the left figure of (g), the optimal training steps for oAUC and cAUC are usually mismatched. Specifically, the line of cAUC (in \textcolor{red}{red}) in the left figure of (g) is usually above the other one (in \textcolor{blue}{blue}), meaning the optimal step for cAUC is usually {\em larger} than oAUC's. This phenomenon reveals that more training steps of the CTR prediction model in a streaming learning scenario usually {\em benefit} the performance on the current distribution and {\em degrade} the performance on the unseen distribution. A possible reason for this trend is that more training steps will {\em overfit} the model to the current distribution and lose the adaptation ability to future distribution.  
    \item {\bf The oAUC usually drops more drastically than cAUC}. As shown in the right figure of (g), the line of oAUC (in \textcolor{blue}{blue}) is usually above the line of cAUC (in \textcolor{red}{red}), meaning the oAUC is more susceptible by the training steps than the cAUC.
    \item {\bf The PD of oAUC tends to {\em saturate}}. As shown in the (a)$\sim$(f) and the right figure of (g), the oAUC in figure (a)$\sim$(f) tends to gradually drop slightly and the PD of oAUC in the right figure of (g) exhibit a slight downward trend.
\end{itemize}
\begin{table*}[]
\small
\centering
\caption{The comparison results of different optimizers. We take the results of the Adam optimizer as the baseline and show the performance improvement (\ie, Imp) of other optimizers. The Adam-based optimizers show superior performance.}
\renewcommand{\arraystretch}{1.1}
\begin{tabular}{l|ll|llllll}
\toprule
 & \multicolumn{2}{c|}{\textbf{CTR (Pre-Train)}} & \multicolumn{6}{c}{\textbf{Streaming CTR}}                    \\ \hline
 &
  \multicolumn{1}{c}{AUC} &
  \multicolumn{1}{c|}{Imp.} &
  \multicolumn{1}{c}{bAUC} &
  \multicolumn{1}{c}{Imp.} &
  \multicolumn{1}{c}{cAUC} &
  \multicolumn{1}{c}{Imp.} &
  \multicolumn{1}{c}{oAUC} &
  \multicolumn{1}{c}{Imp.} \\ \hline
\rowcolor{mygray-bg} Adam &0.7893& - & 0.7936& - &0.7885 & - &0.7691 & - \\
AdamW          & 0.7893& -0.0000& 0.7933& -0.0003 &0.7882 & -0.0003 &0.7687 & -0.0004 \\
\hline
SGD       & 0.7723& -0.0170& 0.7601& -0.0335 &0.7592 & -0.0293 &0.7551 & -0.0140 \\
RMSprop & 0.7865& -0.0028& 0.7938& +0.0002 &0.7921 & +0.0036 &0.7659 & -0.0032  \\ \bottomrule
\end{tabular}\label{tab:optim_analy}
\end{table*}
\subsection{Correlation Analysis}
\label{sec:corr_analy}
\begin{figure}
    \centering
    \includegraphics[width=1.0\linewidth]{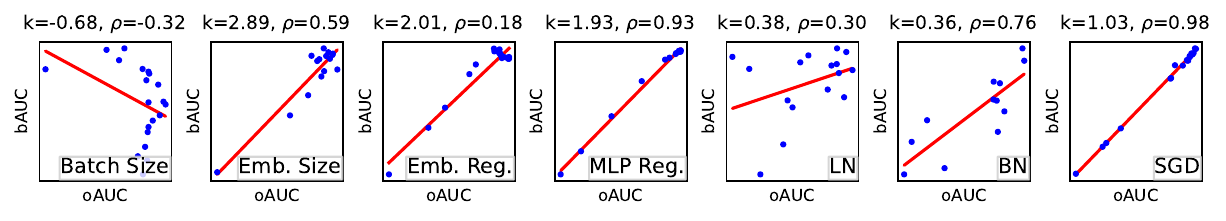}
    \caption{A correlation analysis between the oAUC and bAUC. We fit a linear regression between oAUC and bAUC and show the coefficient with $k$. Besides, we compute the Spearman’s Correlation $\rho$ between oAUC and bAUC. 
    }\label{fig:corr_analysis}
    \vspace{-1mm}
\end{figure}
In previous studies~\cite{congcong_concept_drift,sigir23_asys, forgetting}, the notorious {\em catastrophic forgetting problem} is widely recognized as the primary challenge when updating the CTR prediction model in a streaming context. However, incremental learning focus on acquiring new knowledge while retaining previously learned knowledge (\ie, without forgetting), without the need to infer on unseen distributions. Therefore, the {\em forgetting problem} may not directly influence the model in the streaming CTR prediction task. For example, the well-studied Class-Incremental Learning task~\cite{iCaRL,CIL-survey} in the computer vision field evaluates the model on all {\em seen classes} and does not involve the recognition of {\em unseen new classes}. However, a realistic streaming learning scenario in the recommender system only focuses on the prediction of the {\em unseen new distribution} and does not directly evaluate the model in {\em seen distributions}. Therefore, {\bf restricting the model not to forget is not {\em explicitly} helpful in the Streaming CTR prediction scenario}. To explore whether catastrophic forgetting mitigation can help deep CTR prediction models improve online performance (\ie, oAUC) in a Streaming CTR prediction scenario, we analyze the correlation between the bAUC and oAUC to reveal whether the elimination of forgetting still benefits the model in the streaming CTR prediction scenario. Notably, a larger bAUC indicates less forgetting of the deep CTR prediction model. 

As shown in Figure~\ref{fig:corr_analysis}, we analyze the correlation between oAUC and bAUC under some factors. We first fit a linear regression model between the bAUC and oAUC and show the coefficient $k$ of the regression model. Besides, we also compute Spearman's Correlation between bAUC and oAUC. The majority of findings suggest a notable positive correlation between the oAUC and bAUC metrics, indicating that reducing forgetting can potentially improve the online performance (\ie, oAUC) of deep CTR prediction models in a streaming scenario. Interestingly, the Batch Size results reveal an {\em unexpected negative correlation} between oAUC and bAUC, which suggests that certain factors may challenge conventional wisdom in this context.
\section{Improvement for Streaming CTR Prediction}

\subsection{Methods}
To better understand the online performance of existing CTR prediction models and further improve it, we propose two simple yet effective improvements concerning the online performance (\ie, oAUC).

{\noindent \bf The tuning of key factors concerning the oAUC}. Previous studies on the CTR prediction task lack an understanding of the streaming scenario. Therefore, the hyperparameters of relative optimal performance \wrt static scenario may be sub-optimal to the streaming scenario. Motivated by this {\em ``streaming learning dilemma''}, we tune the existing CTR prediction models under the streaming scenario \wrt the key factors identified from our analysis in section~\ref{sec:main_exp}. Since tuning the key hyperparameters does not incur additional training costs, we refer to this tuning with respect to the key factors as a ``{\em free lunch}'' for the streaming scenario (\ie, the {\bf w/ FL} in Table~\ref{tab:final_performance}).

{\noindent \bf Exemplar replay.} As shown in section~\ref{sec:corr_analy}, the CTR prediction model in a streaming scenario still benefits from mitigating the forgetting phenomenon. To conquer the catastrophic forgetting problem in incremental learning, the most simple yet effective is to store the old data and replay it~\cite{CIL-survey}. In the context of \nname, we improved the existing CTR prediction model and achieved significant enhancement. Notably, the module of saving and replaying exemplars is independent of the CTR prediction models and can be viewed as a plug-and-play optimization for existing models. The results of exemplar replaying are shown as the {\bf w/ Exemplar} in Table~\ref{tab:final_performance}.

\subsection{Results}
We present the comparison results and improvements on vanilla baseline methods in Table~\ref{tab:final_performance}. In addition to the online performance metric oAUC, we also report the bAUC and cAUC, which also exhibit a relative improvement. The relative improvements in bAUC and cAUC suggest that tuning the key factors and exemplar replaying help the model to better adapt to the current distribution and retain the knowledge of the past distributions.

\begin{table*}[t]
\small
\centering
\caption{The comparison results between the baseline methods and their counterparts. {\bf w/ FL} means the baseline methods with tuning key hyperparameters. This improvement for baseline methods can be seen as a {\bf F}ree {\bf L}unch. {\bf w/ Exemplar} means the baseline methods with replayed exemplars. The {\bf Imp.} column exhibits the relative improvement \wrt the vanilla baseline methods.
}
\renewcommand{\arraystretch}{1.1}
\begin{tabular}{l|llllll}
\toprule
\multicolumn{6}{c}{\textbf{Streaming CTR}}                    \\ \hline
 &
  \multicolumn{1}{p{1cm}}{bAUC} &
  \multicolumn{1}{p{1cm}}{Imp.} &
  \multicolumn{1}{p{1cm}}{cAUC} &
  \multicolumn{1}{p{1cm}}{Imp.} &
  \multicolumn{1}{p{1cm}}{oAUC} &
  \multicolumn{1}{p{1cm}}{Imp.} \\ \hline
FM  & 0.7811 & - & 0.7773 & - & 0.7516 & - \\
\rowcolor{mygray-bg} w/ Exemplar  & 0.7832& +0.0021& 0.7782& +0.0009& {\bf 0.7534}& {\bf +0.0018} \\
\hline
xDeepFM &0.7949 & - & 0.7907 & - & 0.7681 & - \\
\rowcolor{mygray-bg} w/ FL &0.7962& +0.0013& 0.7926& +0.0019& \textbf{0.7695}& \textbf{+0.0014} \\
\rowcolor{mygray-bg} w/ Exemplar &0.7962& +0.0013& 0.7925& +0.0018& \textbf{0.7698}& \textbf{+0.0017} \\
\hline
AutoInt & 0.7867 & - & 0.7822 & - & 0.7654 & - \\
\rowcolor{mygray-bg} w/ FL  & 0.7938& +0.0071& 0.7922& +0.0101& \textbf{0.7663}& \textbf{+0.0008} \\
\rowcolor{mygray-bg} w/ Exemplar  &0.7879& +0.0012& 0.7833& +0.0012& {\bf 0.7661}& {\bf +0.0007} \\
\hline
MaskNet & 0.7976 & - & 0.7939 & - & 0.7670 & - \\
\rowcolor{mygray-bg} w/ FL & 0.7993& +0.0018& 0.7974& +0.0035& \textbf{0.7683}& \textbf{+0.0013} \\
\rowcolor{mygray-bg} w/ Exemplar & 0.7979& +0.0003& 0.7943& +0.0004& {\bf 0.7677}& +{\bf 0.0007} \\

\bottomrule
\end{tabular}\label{tab:final_performance}
\end{table*}

\section{Conclusion}
In this study, we formally define the \nnamee task and conduct a systematic analysis of the factors affecting CTR prediction models in a streaming scenario. Our empirical investigation reveals a ``{\em streaming learning dilemma}'' in which certain factors have varying effects on the performance of CTR prediction models in stationary and streaming contexts. Drawing on our empirical analysis, we propose a straightforward performance benchmark and enhance existing baselines by focusing on key factors and exemplar replay. Our work provides a starting point for further exploration of the \nnamee task, and a more sophisticated data replay algorithm could be included in future efforts.

Notably, although we have designed the analysis framework as comprehensively as possible, the diverse structures of CTR prediction models may render some conclusions regarding certain factors {\em inapplicable} to all CTR prediction model structures. Given this challenge, we aim to {\em qualitatively} identify which factors may exhibit differences in the two scenarios and affect the model's performance in the streaming context, rather than {\em quantitatively} providing absolute conclusions. We hope our experiments can contribute to a better understanding of the streaming training process in the recommender system.

{\noindent {\bf Acknowledgment}: This work is partially supported by National Key R\&D Program of China (2022ZD0114805), NSFC (61921006, 62006112, 62250069).
}

\bibliographystyle{plain}
{\small
\bibliography{reference}}

\begin{thebibliography}{10}

\bibitem{vanilla_LN}
Lei~Jimmy Ba, Jamie~Ryan Kiros, and Geoffrey~E. Hinton.
\newblock Layer normalization.
\newblock {\em CoRR}, abs/1607.06450, 2016.

\bibitem{PRML}
Christopher~M. Bishop.
\newblock {\em Pattern Recognition and Machine Learning (Information Science
  and Statistics)}.
\newblock 2006.

\bibitem{streaming-rs}
Shiyu Chang, Yang Zhang, Jiliang Tang, Dawei Yin, Yi~Chang, Mark~A.
  Hasegawa{-}Johnson, and Thomas~S. Huang.
\newblock Streaming recommender systems.
\newblock In {\em WWW}, pages 381--389, 2017.

\bibitem{bias-and-debias-survey}
Jiawei Chen, Hande Dong, Xiang Wang, Fuli Feng, Meng Wang, and Xiangnan He.
\newblock Bias and debias in recommender system: {A} survey and future
  directions.
\newblock {\em CoRR}, abs/2010.03240, 2020.

\bibitem{wide_deeplearning_google}
Heng{-}Tze Cheng, Levent Koc, Jeremiah Harmsen, Tal Shaked, Tushar Chandra,
  Hrishi Aradhye, Glen Anderson, Greg Corrado, Wei Chai, Mustafa Ispir, Rohan
  Anil, Zakaria Haque, Lichan Hong, Vihan Jain, Xiaobing Liu, and Hemal Shah.
\newblock Wide {\&} deep learning for recommender systems.
\newblock In {\em DLRS@RecSys}, pages 7--10, 2016.

\bibitem{DNN_youtube}
Paul Covington, Jay Adams, and Emre Sargin.
\newblock Deep neural networks for youtube recommendations.
\newblock In {\em RecSys}, pages 191--198, 2016.

\bibitem{forgetting}
Robert~M. French and Nick Chater.
\newblock Using noise to compute error surfaces in connectionist networks: {A}
  novel means of reducing catastrophic forgetting.
\newblock {\em Neural Computation}, 14(7):1755--1769, 2002.

\bibitem{gama2010knowledge}
Joao Gama.
\newblock {\em Knowledge discovery from data streams}.
\newblock CRC Press, 2010.

\bibitem{streaming_learning_evaluation}
Jo{\~{a}}o Gama, Raquel Sebasti{\~{a}}o, and Pedro~Pereira Rodrigues.
\newblock On evaluating stream learning algorithms.
\newblock {\em Machine Learning}, 90(3):317--346, 2013.

\bibitem{second-order-gradient}
Fengli Gao and Huicai Zhong.
\newblock Study on the large batch size training of neural networks based on
  the second order gradient.
\newblock {\em CoRR}, abs/2012.08795, 2020.

\bibitem{DeepFM}
Huifeng Guo, Ruiming Tang, Yunming Ye, Zhenguo Li, and Xiuqiang He.
\newblock Deepfm: {A} factorization-machine based neural network for {CTR}
  prediction.
\newblock In {\em IJCAI}, pages 1725--1731, 2017.

\bibitem{ESL}
Trevor Hastie, Robert Tibshirani, and Jerome Friedman.
\newblock {\em The Elements of Statistical Learning}.
\newblock 2001.

\bibitem{evolvable-stream-aaai2021}
Bo{-}Jian Hou, Yu{-}Hu Yan, Peng Zhao, and Zhi{-}Hua Zhou.
\newblock Storage fit learning with feature evolvable streams.
\newblock In {\em AAAI}, pages 7729--7736, 2021.

\bibitem{feature-evolvable-stream-nips2017}
Bo{-}Jian Hou, Lijun Zhang, and Zhi{-}Hua Zhou.
\newblock Learning with feature evolvable streams.
\newblock In {\em NIPS}, pages 1417--1427, 2017.

\bibitem{hou2019learning}
Saihui Hou, Xinyu Pan, Chen~Change Loy, Zilei Wang, and Dahua Lin.
\newblock Learning a unified classifier incrementally via rebalancing.
\newblock In {\em CVPR}, pages 831--839, 2019.

\bibitem{vanilla_BN}
Sergey Ioffe and Christian Szegedy.
\newblock Batch normalization: Accelerating deep network training by reducing
  internal covariate shift.
\newblock In {\em ICML}, pages 448--456, 2015.

\bibitem{DBLP:journals/pami/JosephRKKB22}
K.~J. Joseph, Jathushan Rajasegaran, Salman~H. Khan, Fahad~Shahbaz Khan, and
  Vineeth~N. Balasubramanian.
\newblock Incremental object detection via meta-learning.
\newblock {\em IEEE Transactions on Pattern Analysis and Machine Intelligence},
  44(12):9209--9216, 2022.

\bibitem{large-batch-training}
Nitish~Shirish Keskar, Dheevatsa Mudigere, Jorge Nocedal, Mikhail Smelyanskiy,
  and Ping Tak~Peter Tang.
\newblock On large-batch training for deep learning: Generalization gap and
  sharp minima.
\newblock In {\em ICLR}, 2017.

\bibitem{adam}
Diederik~P. Kingma and Jimmy Ba.
\newblock Adam: {A} method for stochastic optimization.
\newblock In {\em ICLR}, 2015.

\bibitem{Ensemble_and_streaming_learning}
Bartosz Krawczyk, Leandro~L. Minku, Jo{\~{a}}o Gama, Jerzy Stefanowski, and
  Michal Wozniak.
\newblock Ensemble learning for data stream analysis: {A} survey.
\newblock {\em Information Fusion}, 37:132--156, 2017.

\bibitem{delayed-feedback-for-continuous-training}
Sofia~Ira Ktena, Alykhan Tejani, Lucas Theis, Pranay~Kumar Myana, Deepak
  Dilipkumar, Ferenc Husz{\'{a}}r, Steven Yoo, and Wenzhe Shi.
\newblock Addressing delayed feedback for continuous training with neural
  networks in {CTR} prediction.
\newblock In {\em RecSys}, pages 187--195, 2019.

\bibitem{AdaBN}
Yanghao Li, Naiyan Wang, Jianping Shi, Xiaodi Hou, and Jiaying Liu.
\newblock Adaptive batch normalization for practical domain adaptation.
\newblock {\em Pattern Recognition}, 80:109--117, 2018.

\bibitem{xdeepfm}
Jianxun Lian, Xiaohuan Zhou, Fuzheng Zhang, Zhongxia Chen, Xing Xie, and
  Guangzhong Sun.
\newblock xdeepfm: Combining explicit and implicit feature interactions for
  recommender systems.
\newblock In {\em KDD}, pages 1754--1763, 2018.

\bibitem{TTA_survey}
Jian Liang, Ran He, and Tieniu Tan.
\newblock A comprehensive survey on test-time adaptation under distribution
  shifts.
\newblock {\em CoRR}, abs/2303.15361, 2023.

\bibitem{congcong_concept_drift}
Congcong Liu, Yuejiang Li, Xiwei Zhao, Changping Peng, Zhangang Lin, and
  Jingping Shao.
\newblock Concept drift adaptation for {CTR} prediction in online advertising
  systems.
\newblock {\em CoRR}, abs/2204.05101, 2022.

\bibitem{sigir23_asys}
Congcong Liu, Fei Teng, Xiwei Zhao, Zhangang Lin, Jinghe Hu, and Jingping Shao.
\newblock Always strengthen your strengths: {A} drift-aware incremental
  learning framework for {CTR} prediction.
\newblock {\em CoRR}, abs/2304.09062, 2023.

\bibitem{previous-model-bias}
David~C. Liu, Stephanie~Kaye Rogers, Raymond Shiau, Dmitry Kislyuk, Kevin~C.
  Ma, Zhigang Zhong, Jenny Liu, and Yushi Jing.
\newblock Related pins at pinterest: The evolution of a real-world recommender
  system.
\newblock In {\em WWW}, pages 583--592, 2017.

\bibitem{rmm}
Yaoyao Liu, Bernt Schiele, and Qianru Sun.
\newblock {RMM:} reinforced memory management for class-incremental learning.
\newblock In {\em NeurIPS}, pages 3478--3490, 2021.

\bibitem{concurrent-adversarial-learning-for-large-batch-training}
Yong Liu, Xiangning Chen, Minhao Cheng, Cho{-}Jui Hsieh, and Yang You.
\newblock Concurrent adversarial learning for large-batch training.
\newblock In {\em ICLR}, 2022.

\bibitem{learning-under-concept-drift}
Jie Lu, Anjin Liu, Fan Dong, Feng Gu, Jo{\~{a}}o Gama, and Guangquan Zhang.
\newblock Learning under concept drift: {A} review.
\newblock {\em IEEE Transactions on Knowledge and Data Engineering},
  31(12):2346--2363, 2019.

\bibitem{session_based_memory_augmented_inc}
Fei Mi and Boi Faltings.
\newblock Memory augmented neural model for incremental session-based
  recommendation.
\newblock In {\em IJCAI}, pages 2169--2176, 2020.

\bibitem{session_based_exemplar_ader}
Fei Mi, Xiaoyu Lin, and Boi Faltings.
\newblock {ADER:} adaptively distilled exemplar replay towards continual
  learning for session-based recommendation.
\newblock In {\em RecSys}, pages 408--413, 2020.

\bibitem{iCaRL}
Sylvestre{-}Alvise Rebuffi, Alexander Kolesnikov, Georg Sperl, and Christoph~H.
  Lampert.
\newblock icarl: Incremental classifier and representation learning.
\newblock In {\em CVPR}, pages 5533--5542, 2017.

\bibitem{fm}
Steffen Rendle.
\newblock Factorization machines.
\newblock In {\em ICDM}, pages 995--1000, 2010.

\bibitem{l2_regularization}
J{\"{u}}rgen Schmidhuber.
\newblock Deep learning in neural networks: An overview.
\newblock {\em Neural Networks}, 61:85--117, 2015.

\bibitem{deepcrossing}
Ying Shan, T.~Ryan Hoens, Jian Jiao, Haijing Wang, Dong Yu, and J.~C. Mao.
\newblock Deep crossing: Web-scale modeling without manually crafted
  combinatorial features.
\newblock In {\em KDD}, pages 255--262, 2016.

\bibitem{incre-batch-size}
Samuel~L. Smith, Pieter{-}Jan Kindermans, Chris Ying, and Quoc~V. Le.
\newblock Don't decay the learning rate, increase the batch size.
\newblock In {\em ICLR}, 2018.

\bibitem{autoint}
Weiping Song, Chence Shi, Zhiping Xiao, Zhijian Duan, Yewen Xu, Ming Zhang, and
  Jian Tang.
\newblock Autoint: Automatic feature interaction learning via self-attentive
  neural networks.
\newblock In {\em CIKM}, pages 1161--1170, 2019.

\bibitem{dropout}
Nitish Srivastava, Geoffrey~E. Hinton, Alex Krizhevsky, Ilya Sutskever, and
  Ruslan Salakhutdinov.
\newblock Dropout: a simple way to prevent neural networks from overfitting.
\newblock {\em Journal of Machine Learning Research}, 15(1):1929--1958, 2014.

\bibitem{four-things-should-know-bn}
Cecilia Summers and Michael~J. Dinneen.
\newblock Four things everyone should know to improve batch normalization.
\newblock In {\em ICLR}, 2020.

\bibitem{the-problem-of-concept-drift}
Alexey Tsymbal.
\newblock The problem of concept drift: definitions and related work.
\newblock {\em Computer Science Department, Trinity College Dublin}, 106(2):58,
  2004.

\bibitem{streaming_learning_RS_evaluation}
Jo{\~{a}}o Vinagre, Al{\'{\i}}pio~M{\'{a}}rio Jorge, and Jo{\~{a}}o Gama.
\newblock Evaluation of recommender systems in streaming environments.
\newblock {\em CoRR}, abs/1504.08175, 2015.

\bibitem{DBLP:conf/iccv/WangWSG21}
Jianren Wang, Xin Wang, Yue Shang{-}Guan, and Abhinav Gupta.
\newblock Wanderlust: Online continual object detection in the real world.
\newblock In {\em ICCV}, pages 10809--10818, 2021.

\bibitem{DCN}
Ruoxi Wang, Bin Fu, Gang Fu, and Mingliang Wang.
\newblock Deep {\&} cross network for ad click predictions.
\newblock In {\em ADKDD}, pages 12:1--12:7, 2017.

\bibitem{incCTR}
Yichao Wang, Huifeng Guo, Ruiming Tang, Zhirong Liu, and Xiuqiang He.
\newblock A practical incremental method to train deep {CTR} models.
\newblock {\em CoRR}, abs/2009.02147, 2020.

\bibitem{MaskNet}
Zhiqiang Wang, Qingyun She, and Junlin Zhang.
\newblock Masknet: Introducing feature-wise multiplication to {CTR} ranking
  models by instance-guided mask.
\newblock {\em CoRR}, abs/2102.07619, 2021.

\bibitem{correct_norm}
Zhiqiang Wang, Qingyun She, Pengtao Zhang, and Junlin Zhang.
\newblock Correct normalization matters: Understanding the effect of
  normalization on deep neural network models for click-through rate
  prediction.
\newblock {\em CoRR}, abs/2006.12753, 2020.

\bibitem{unbiased-learning-to-ran-feed-recommendation}
Xinwei Wu, Hechang Chen, Jiashu Zhao, Li~He, Dawei Yin, and Yi~Chang.
\newblock Unbiased learning to rank in feeds recommendation.
\newblock In {\em WSDM}, pages 490--498, 2021.

\bibitem{rethinking_batch}
Yuxin Wu and Justin Johnson.
\newblock Rethinking "batch" in batchnorm.
\newblock {\em CoRR}, abs/2105.07576, 2021.

\bibitem{simpleLN}
Jingjing Xu, Xu~Sun, Zhiyuan Zhang, Guangxiang Zhao, and Junyang Lin.
\newblock Understanding and improving layer normalization.
\newblock In {\em NeurIPS}, pages 4383--4393, 2019.

\bibitem{generalized-delay-feedback}
Jia{-}Qi Yang and De{-}Chuan Zhan.
\newblock Generalized delayed feedback model with post-click information in
  recommender systems.
\newblock In {\em NeurIPS}, 2022.

\bibitem{unbiased-position-aware}
Bo{-}Wen Yuan, Yaxu Liu, Jui{-}Yang Hsia, Zhenhua Dong, and Chih{-}Jen Lin.
\newblock Unbiased ad click prediction for position-aware advertising systems.
\newblock In {\em RecSys}, pages 368--377, 2020.

\bibitem{CTR_survey}
Weinan Zhang, Jiarui Qin, Wei Guo, Ruiming Tang, and Xiuqiang He.
\newblock Deep learning for click-through rate estimation.
\newblock In {\em IJCAI}, pages 4695--4703, 2021.

\bibitem{oneEpoch}
Zhao{-}Yu Zhang, Xiang{-}Rong Sheng, Yujing Zhang, Biye Jiang, Shuguang Han,
  Hongbo Deng, and Bo~Zheng.
\newblock Towards understanding the overfitting phenomenon of deep
  click-through rate prediction models.
\newblock {\em CoRR}, abs/2209.06053, 2022.

\bibitem{evolvable-stream-icml2020}
Zhenyu Zhang, Peng Zhao, Yuan Jiang, and Zhi{-}Hua Zhou.
\newblock Learning with feature and distribution evolvable streams.
\newblock In {\em ICML}, pages 11317--11327, 2020.

\bibitem{CIL-survey}
Da{-}Wei Zhou, Qi{-}Wei Wang, Zhi{-}Hong Qi, Han{-}Jia Ye, De{-}Chuan Zhan, and
  Ziwei Liu.
\newblock Deep class-incremental learning: {A} survey.
\newblock {\em CoRR}, abs/2302.03648, 2023.

\bibitem{DIN}
Guorui Zhou, Xiaoqiang Zhu, Chengru Song, Ying Fan, Han Zhu, Xiao Ma, Yanghui
  Yan, Junqi Jin, Han Li, and Kun Gai.
\newblock Deep interest network for click-through rate prediction.
\newblock In {\em KDD}, pages 1059--1068, 2018.

\bibitem{BARS}
Jieming Zhu, Quanyu Dai, Liangcai Su, Rong Ma, Jinyang Liu, Guohao Cai,
  Xi~Xiao, and Rui Zhang.
\newblock {BARS:} towards open benchmarking for recommender systems.
\newblock In {\em SIGIR}, pages 2912--2923, 2022.

\bibitem{BarsCTR}
Jieming Zhu, Jinyang Liu, Shuai Yang, Qi~Zhang, and Xiuqiang He.
\newblock Open benchmarking for click-through rate prediction.
\newblock In {\em CIKM}, pages 2759--2769, 2021.

\end{thebibliography}

\newpage
\appendix

\end{document}